\documentclass[journal]{IEEEtran}

\usepackage{amssymb}
\usepackage[cmex10]{amsmath}
\usepackage{amsfonts}
\usepackage{lineno}
\modulolinenumbers[5]
\usepackage{textgreek}
\usepackage{tikz}
\usetikzlibrary{arrows.meta}
\usepackage{graphicx}
\usepackage{grffile}
\graphicspath{{./fig/},{./photos/}}
\DeclareGraphicsExtensions{.pdf,.jpeg,.jpg,.png}
\usepackage{xcolor}
\usepackage{array}
\usepackage{multirow}
\usepackage{tabulary}
\usepackage{booktabs}
\usepackage{setspace}
\usepackage{algorithm}
\usepackage{algorithmicx}
\usepackage{algpseudocode}

\newcommand{\isot}[2]{\(^{\mathrm{#2}}\){#1}}

\DeclareMathOperator*{\diag}{diag}
\DeclareMathOperator*{\trace}{tr}

\usepackage{soul}
\usepackage{etoolbox}
\makeatletter
\patchcmd{\SOUL@ulunderline}{\dimen@}{\SOUL@dimen}{}{}
\patchcmd{\SOUL@ulunderline}{\dimen@}{\SOUL@dimen}{}{}
\patchcmd{\SOUL@ulunderline}{\dimen@}{\SOUL@dimen}{}{}
\newdimen\SOUL@dimen
\makeatother

\usepackage{url}
\usepackage{hyperref}
\usepackage[plain]{fancyref}

\begin{document}
\IEEEpubid{\begin{minipage}{0.85\textwidth}\ \\[12pt]\\ \\ \\ \centering
  \textcopyright 2021 IEEE.  Personal use of this material is permitted. Permission from IEEE must be obtained for all other uses, in any current or future media, including reprinting/republishing this material for advertising or promotional purposes, creating new collective works, for resale or redistribution to servers or lists, or reuse of any copyrighted component of this work in other works.
\end{minipage}}

\title{Improved Gamma-Ray Point Source Quantification in Three Dimensions by Modeling Attenuation in the Scene}

\author{M.\,S.~Bandstra,
    D.~Hellfeld,
    J.\,R.~Vavrek,
    B.\,J.~Quiter,
    K.~Meehan,
    P.\,J.~Barton,
    J.\,W.~Cates,
    A.~Moran,
    V.~Negut,
    R.~Pavlovsky,
    and T.\,H.\,Y.~Joshi

\thanks{This material is based upon work supported by the Defense Threat Reduction Agency under HDTRA 10027--28018, 10027--30529.
This support does not constitute an express or implied endorsement on the part of the United States Government.
Distribution A: approved for public release, distribution is unlimited.}
\thanks{All authors are with the Nuclear Science Division at Lawrence Berkeley National Laboratory, Berkeley, CA 94720 USA (e-mail: msbandstra@lbl.gov).}%
}

\maketitle

\pagenumbering{gobble}

\begin{abstract}
Using a series of detector measurements taken at different locations to localize a source of radiation is a well-studied problem.
The source of radiation is sometimes constrained to a single point-like source, in which case the location of the point source can be found using techniques such as maximum likelihood.
Recent advancements have shown the ability to locate point sources in 2D and even 3D, but few have studied the effect of intervening material on the problem.
In this work we examine gamma-ray data taken from a freely moving system and develop voxelized 3-D models of the scene using data from the onboard LiDAR\@.
Ray casting is used to compute the distance each gamma ray travels through the scene material, which is then used to calculate attenuation assuming a single attenuation coefficient for solids within the geometry.
Parameter estimation using maximum likelihood is performed to simultaneously find the attenuation coefficient, source activity, and source position that best match the data.
Using a simulation, we validate the ability of this method to reconstruct the true location and activity of a source, along with the true attenuation coefficient of the structure it is inside, and then we apply the method to measured data with sources and find good agreement.
\end{abstract}

\IEEEpeerreviewmaketitle%



\section{Introduction}
\IEEEPARstart{S}{earching} for, localizing, and quantifying radioactive material using radiation detectors is a problem with a wide variety of applications, from finding lost sources, to mapping radioactive contamination after a nuclear accident, to stopping the transport of material at a border crossing~\cite{kouzes_detecting_2005, runkle_photon_2009}.
In many applications, the presence of passive material between the sources and the instrument, if not properly accounted for, can limit the ability of an operator to quantify the source and, e.g., lead to an underestimation or erroneous localization of the radioactivity.
Often in these situations the attenuating material has three-dimensional structure, which adds another degree of complexity.
Because of the difficulty in sensing and quantifying the attenuation and scattering by passive material, these effects are seldom accounted for in search applications, but nevertheless they can be of vital importance.

Many nuclear security, safety and non-proliferation applications using radiation detection and/or imaging systems can benefit from increased abilities to quantify sources in three-dimensional environments with attenuation from material in those environments.
For example, better mapping of radioactive sources in three-dimensional volumes can help account for material stored in casks or waste drums in nuclear safeguards applications.
In the event of a nuclear accident, the ability to rapidly map and quantify contamination, especially in complex areas like urban environments where the shielding from structures is important, could provide responders and the public with high-quality, actionable information needed to reduce their dose.

In some gamma-ray imaging applications where the quantification of radiological material is critical, techniques have been developed to simultaneously solve for the radiation source distribution and attenuation by passive material.
Examples are in Positron Emission Tomography (PET) and Single Photon Emission Computed Tomography (SPECT) imaging, where algorithms have been developed to simultaneously image the target while optimizing the unknown attenuation due to material in the body~\cite{dicken_new_1999, nuyts_simultaneous_1999, gourion_attenuation_2002}.
Such techniques have been found to improve upon earlier techniques that used X-ray computed tomography (CT) to estimate the attenuation map~\cite{kinahan_x-ray-based_2003}.
For a recent review of this field, see~\cite{berker_attenuation_2016}.
Similar techniques have been developed for Passive Gamma Emission Tomography (PGET), where tomographic measurements are used to inspect spent nuclear fuel assemblies~\cite{backholm_simultaneous_2020}.
The optimizations required are nonconvex~\cite{berker_attenuation_2016}, so these techniques generally leverage the constrained detector-image  geometry, a dense image space with a limited spatial extent that the detectors encircle, and predictable attenuation coefficients (e.g., human tissue or fuel pins).
Generally these constraints are included in the form of regularization functions that are added to the loss function.
Some techniques even incorporate non-attenuation contextual data such as magnetic resonance imaging (MRI) to aid in the reconstruction~\cite{salomon_simultaneous_2011}.

However, in the case of a freely moving detector system, none of these advantages remain.
No longer is the space constrained, nor can one assume much if anything about the density of material in the scene, or even if the source is in the same plane as the detectors.
Nevertheless, many have studied the problem of using a series of detector measurements to locate and measure the strength of a radiation point source, using a variety of different techniques, including maximum likelihood estimation~\cite{morelande_detection_2007, vilim_integrated_2011, wan_detection_2012, deb_iterative_2013, bai_maximum_2015, hellfeld_gamma-ray_2019}, Bayesian estimation~\cite{morelande_detection_2007, towler_radiation_2012, miller_adaptively_2015, hite_localization_2019, anderson_mobile_2020}, geometric techniques~\cite{chin_identification_2010, xu_computational_2010}, particle filters~\cite{rao_network_2015}, and clustering techniques~\cite{wu_network_2019}.
Solutions to this problem have been demonstrated in 2D~\cite{morelande_detection_2007, vilim_integrated_2011, towler_radiation_2012, deb_iterative_2013, bai_maximum_2015, vilim_radtrac:_2009, klann_treatment_2011}, and often a grid is used to simplify the calculations involved and to fully explore the nonconvex space~\cite{towler_radiation_2012, miller_adaptively_2015, cordone_improved_2017}.
Because of the computational complexity and the difficulty of obtaining a representative model of the measurement geometry and incorporating it into an algorithm, attenuation by intervening material is seldom considered~\cite{vilim_radtrac:_2009, klann_treatment_2011, hite_localization_2019, anderson_mobile_2020}.
Only recently have source localization methods been applied in 3D without attenuation~\cite{sharma_three-dimensional_2016, bhattacharyya_estimating_2018}.
In one recent work, 2-D point source localization was performed with precomputed attenuation coefficients to account for objects in the scene~\cite{hite_localization_2019}, and in another recent work 3-D point source localization was performed with unknown attenuation between the source and detector but the thicknesses and orientations of the attenuating objects were known~\cite{anderson_mobile_2020}.
Our recent work has demonstrated accurate quantitative reconstruction in 3D with freely moving systems in cases when attenuation can be ignored~\cite{hellfeld_gamma-ray_2019, vavrek_reconstructing_2020, hellfeld_free-moving_2021}.
To our knowledge, the method has not been applied both in 3D and with attenuation by unknown objects.

As challenging as the SPECT and PGET reconstruction problems are, the problem of source reconstruction in 3D with a freely moving system has even fewer constraints, even before the simultaneous reconstruction of attenuation is considered.
To make the problem tractable, we make four simplifying assumptions herein: (1) that we are searching for a single point source, (2) that the background is relatively constant, (3) that the contextual data from the LiDAR point cloud constrains where solid material in the scene can be, and (4) that all solid materials have the same attenuation coefficient.
The objective of this work is to develop a method to solve this special case and to demonstrate the proof-of-concept of the approach using measured data.

The solution framework we will use is referred to as Point Source Likelihood (PSL)~\cite{hellfeld_gamma-ray_2019, vavrek_reconstructing_2020, hellfeld_free-moving_2021}.
PSL requires the knowledge of a detector's position and orientation in 3D as it collects gamma-ray events, as well as knowledge of the detector's angular response.
The goal of PSL is to exploit the full 3-D trajectory and orientation of a freely moving detector system, together with its (often sparse) gamma-ray event data, and infer the position and activity of point sources, including uncertainties on all quantities.
To achieve this goal, PSL uses a sparse representation of the problem, namely that a single point source and constant background give rise to the detector measurements, and it tests for the possible presence of a point source at a number of fixed positions in space.
PSL has also been implemented with a continuous solver instead of using discrete points~\cite{hellfeld_gamma-ray_2019}, and it has been extended to successfully solve for multiple point sources~\cite{vavrek_reconstructing_2020}.

In this paper we will show how PSL can be extended to include the attenuation of material in the scene (\Fref{sec:methods}) and demonstrate with a toy model that fitting the attenuation parameter can return the correct result (\Fref{sec:toy_model}).
We then apply the method to measurements of sources in experimental scenarios (\Fref{sec:results}) and conclude by discussing the effectiveness and limitations of the approach (\Fref{sec:discussion}).


\section{Methodology}\label{sec:methods}
In this section, we will explain the existing PSL algorithm, and then extend and validate a version that includes attenuation in the scene.


\subsection{PSL without attenuation}
In the existing PSL approach, the 3-D environment around the detection system is first populated with possible positions of point sources (test points).
For example, a regular grid of 3-D space around the detector's path may be used, and that grid may potentially be downselected to consider only those grid points that are near a surface (inferred from, e.g., a point cloud).
The photon events in question are assumed to come either from a photopeak or from a background that is constant in time; downscattering is neglected.
A response matrix is generated that gives the average number of photopeak photons that would be detected in measurement \(i\) for a source of unit activity at test point \(j\).
The response matrix \(\mathbf{R}\), which relates source activity to detector count rates, is calculated using the formula
\begin{align}
R_{ij} &= \frac{A(\mathbf{q}_i, \hat{\mathbf{r}}_{ij})}{4 \pi |\mathbf{r}_{ij}|^2} B C \tau_{ij},
\end{align}
where \(\mathbf{q}_i\) is the system's orientation, \(\mathbf{r}_{ij}\) is the vector from the system to the test point, \(A(\mathbf{q}_i, \hat{\mathbf{r}}_{ij})\) is the effective area (geometric area times efficiency) given the system orientation and direction to the point, which we have assumed is in the far field, \(B\) is the branching ratio for the gamma-ray emission of interest, \(C\) is a conversion factor to relate the source activity units to the number of nuclear decays per second, and \(\tau_{ij}\) is the fraction of unattenuated photopeak photons along \(\mathbf{r}_{ij}\).
For the case of no attenuation, we will assume \(\tau_{ij} = 1\).

For each test point \(j\), a linear model is considered that assumes all detections are either from a point source at \(j\) or a constant background:
\begin{align}
n_i &\sim \mathrm{Poisson}(\mu_{ij}), \label{eq:ni} \\
\mu_{ij} &\equiv (R_{ij} s + b) \Delta t_i, \label{eq:muij}
\end{align}
where \(s\) is the activity of the source at the test point, \(b\) is the background count rate, and \(\Delta t_i\) is the integration time.
To simplify the description of the linear model, we introduce the system tensor
\begin{align}
    X_{ij0} &= R_{ij} \Delta t_i \\
    X_{ij1} &= \Delta t_i
\end{align}
and then shift notation by vectorizing over the measurement index \(i\) and using boldface for the resulting vectors; e.g., the measured photopeak counts \(n_i\) become \(\mathbf{n}\), the mean counts \(\mu_{ij}\) become \(\boldsymbol\mu_j \), and the \(j\)th column of \(R_{ij}\) is \(\mathbf{R}_j\).
In this notation, \(X\) is considered to be a set of 2-D system matrices (dimensions are the number of measurements \(\times\) 2) for each test point \(j\):
\begin{align}
    \mathbf{X}_j = \left[ \mathbf{R}_j \odot \mathbf{\Delta t},\ \mathbf{\Delta t} \right],
\end{align}
where \(\odot\) is element-wise multiplication.
Grouping \(s\) and \(b\) into the parameter vector \(\boldsymbol\theta^{\top} = \left[ s, \ b \right]^{\top}\), the problem solved at each test point \(j\) (\fref{eq:ni} and \fref{eq:muij}) can finally be rewritten as the following matrix equation:
\begin{align}
\mathbf{n} &\sim \mathrm{Poisson}(\boldsymbol\mu_j = \mathbf{X}_j \boldsymbol\theta). \label{eq:poisson}
\end{align}
In accordance with~\Fref{eq:poisson}, the optimal value of \(\boldsymbol\theta\) is found by minimizing the negative Poisson log likelihood function:
\begin{align}
-\log L_j(\boldsymbol\theta) &= \sum_i \left[\mu_{ij} - n_i \log \mu_{ij} + \log (n_i!)\right] \\
&= \sum_i \left[(\mathbf{X}_j \boldsymbol\theta)_i - n_i \log (\mathbf{X}_j \boldsymbol\theta)_i + \log (n_i!)\right] \label{eq:nll}.
\end{align}

The negative log likelihood function \(-\log L_j(\boldsymbol\theta)\) is convex in \(\boldsymbol\theta\)~\cite{shepp_maximum_1982}, which means it has a single, global maximum, and this property can be quickly established by examining its Hessian:
\begin{align}
    \mathbf{H}_j(\boldsymbol\theta) &= \left[\begin{array}{cc}
        \sum_i \frac{R_{ij}^2 \Delta t_i^2 n_i}{\mu_{ij}^2} & \sum_i \frac{R_{ij} \Delta t_i^2 n_i}{\mu_{ij}^2} \\
        \sum_i \frac{R_{ij} \Delta t_i^2 n_i}{\mu_{ij}^2} & \sum_i \frac{\Delta t_i^2 n_i}{\mu_{ij}^2}
    \end{array}\right] \\
    &= \mathbf{X}_j^{\top} \diag \left(\frac{\mathbf{n}}{(\mathbf{X}_j \boldsymbol\theta)^2}\right) \mathbf{X}_j.
\end{align}
By defining
\begin{align}
    \mathbf{v}_j^s \equiv \frac{\mathbf{R}_j \odot \mathbf{\Delta t} \odot \sqrt{\mathbf{n}}}{\mathbf{X}_j \boldsymbol\theta}, & \ \ \
    \mathbf{v}_j^b \equiv \frac{\mathbf{\Delta t} \odot \sqrt{\mathbf{n}}}{\mathbf{X}_j \boldsymbol\theta},
\end{align}
where the square root and division are performed element-wise, we can write the Hessian in the form
\begin{align}
    \mathbf{H}_j &= \left[\begin{array}{cc}
        \mathbf{v}_{j}^s \cdot \mathbf{v}_{j}^s & \mathbf{v}_{j}^s \cdot \mathbf{v}_{j}^b \\
        \mathbf{v}_{j}^s \cdot \mathbf{v}_{j}^b & \mathbf{v}_{j}^b \cdot \mathbf{v}_{j}^b
    \end{array}\right].
\end{align}
And therefore
\begin{align}
    \trace \mathbf{H}_j &= \| \mathbf{v}_{j}^s \|^2 + \| \mathbf{v}_{j}^b \|^2 \ge 0 \\
    \det \mathbf{H}_j &= \| \mathbf{v}_{j}^s \|^2 \| \mathbf{v}_{j}^b \|^2 - \left( \mathbf{v}_{j}^s \cdot \mathbf{v}_{j}^b \right)^2 \ge 0,
\end{align}
where the second inequality follows from the Cauchy-Schwarz inequality.
Because \(\mathbf{H}_j\) is a symmetric \(2 \times 2\) matrix and its trace and determinant are both non-negative, it follows that both of its eigenvalues are non-negative, making \(\mathbf{H}_j\) positive semi-definite, and thus \(-\log L_j\) is everywhere a convex function of \(\boldsymbol\theta\).
Therefore \(-\log L_j\) has a single, global maximum at a point that we will denote \(\hat{\boldsymbol\theta}_j\).
(The only possibility of a degenerate solution is if \(n_i = 0\ \forall i\), a case we avoid by the implicit assumption there is at least one measured event.)

To find \(\hat{\boldsymbol\theta}_j\) for each test point \(j\), the multiplicative Maximum Likelihood Expectation Maximization (MLEM) update rules for Poisson data were used~\cite{shepp_maximum_1982}.
Starting with any initial non-zero guess for \(\hat{\boldsymbol\theta}_j\), the multiplicative update rule is
\begin{align}
\hat{\boldsymbol\theta}_j &\leftarrow \hat{\boldsymbol\theta}_j \odot \left( \frac{\mathbf{X}_j^{\top} \cdot \frac{\mathbf{n}}{\mathbf{X}_j \hat{\boldsymbol\theta}_j}}{\mathbf{X}_j^{\top} \cdot \mathbf{1}_M} \right),
\end{align}
where \(M\) is the number of measurements and \(\mathbf{1}_M\) is a column vector of \(M\) ones.
For each test point, at least 100~iterations of the multiplicative update rules were performed, and iterations were terminated once the change in all parameters was less than 10\(^{-5}\).
Then the negative log likelihood for each test point was calculated was calculated at the best fit parameters:
\begin{align}
-\log \hat{L}_j &\equiv -\log L_j(\hat{\boldsymbol\theta}_j).
\end{align}
The overall most likely location for the point source was determined by finding the minimum negative log likelihood, \(-\log \hat{L}_{\mathrm{min}}\), and its corresponding test point \(j_{\mathrm{min}}\).
We note that although the optimization to find \(\hat{\boldsymbol\theta}_j\) is convex and therefore a global maximum exists, the optimization over \(j\) (i.e., over the three spatial dimensions) is not guaranteed to be convex.

Spatial confidence intervals were determined by using the likelihood ratio test (LRT), where the null hypothesis is the best fit model with all parameters held constant.
The LRT statistic \(\Lambda_j = 2 (-\log \hat{L}_j + \log \hat{L}_{\mathrm{min}})\) is approximately distributed as a chi-squared distribution with five degrees of freedom: one for \(s\), one for \(b\), and three degrees of freedom for the three spatial dimensions, even though they are not explicit model parameters~\cite{wilks_large-sample_1938}.
The necessity of including these three implicit spatial degrees of freedom to correctly estimate confidence intervals was confirmed through simulations.
Choosing an \(\alpha\) to construct a \(1-\alpha\) confidence interval (e.g., \(\alpha = 0.05\) for a 95\% confidence interval) yields the following LRT threshold:
\begin{align}
\Lambda_{\mathrm{threshold}}(\alpha) &= \Phi^{-1}_{\chi^2_5} \left(1 - \alpha\right), \label{eq:lambda_threshold_a}
\end{align}
where \(\Phi^{-1}_{\chi^2_5}\) is the inverse cumulative distribution function of the \(\chi^2\) distribution with five degrees of freedom.
Parametrizing confidence intervals instead using the \(z\)-sigma unit normal equivalent yields:
\begin{align}
\alpha(z) &= 2 \left(1 - \Phi_{\mathrm{normal}}(z) \right) \\
&= 1 - \mathrm{erf}\left(\frac{z}{\sqrt{2}}\right) \\
\Lambda_{\mathrm{threshold}}(z) &= \Phi^{-1}_{\chi^2_5} \left[\mathrm{erf}\left(\frac{z}{\sqrt{2}}\right)\right], \label{eq:lambda_threshold}
\end{align}
where \(\Phi_{\mathrm{normal}}\) is the cumulative distribution function of the unit normal distribution.
All of the test points whose \(\Lambda_j\) values are below the chosen threshold are included in the \(1 - \alpha\) or \(z\)-sigma confidence interval.

Source activity confidence intervals were performed in a different manner.
The source activity interval could be estimated as the interval that encloses the \(\hat{s}_j\) values of all test points within the spatial confidence interval.
However, in practice, especially if the number of such points is low, this method might not be conservative enough.
To compensate for this problem, a confidence interval for \(\hat{s}_j\) was developed for each test point, and the union of these confidence intervals was used as the overall confidence interval for \(s\).
To estimate the confidence interval for \(\hat{s}_j\) at test point \(j\), the observed Fisher information matrix was used, which is just the Hessian of the negative log likelihood evaluated at \(\hat{\boldsymbol\theta}_j\):
\begin{align}
\mathbf{F}_j &= \mathbf{X}_j^{\top} \mathrm{diag}\left(\frac{\mathbf{n}}{(\mathbf{X}_j \hat{\boldsymbol\theta}_j)^2}\right) \mathbf{X}_j.
\end{align}
Holding the test point position constant but considering hypothetical statistical fluctuations (e.g., performing a parametric bootstrap), the negative log likelihood will vary in the neighborhood of \(\hat{\boldsymbol\theta}_j\) according to its Taylor expansion (where the first derivative vanishes by the definition of maximum likelihood):
\begin{align}
-\log L^*_j &\approx -\log \hat{L}_j + \frac{1}{2} (\boldsymbol\theta_j^* - \hat{\boldsymbol\theta}_j)^{\top} \mathbf{F}_j (\boldsymbol\theta_j^* - \hat{\boldsymbol\theta}_j)
\end{align}
We then select the range of \(\boldsymbol\theta_j^*\) values that result in a negative log likelihood below the threshold:
\begin{align}
\Lambda_j + (\boldsymbol\theta_j^* - \hat{\boldsymbol\theta}_j)^{\top} \mathbf{F}_j (\boldsymbol\theta_j^* - \hat{\boldsymbol\theta}_j) &\le \Lambda_{\mathrm{threshold}}(z).
\end{align}
When the above condition reaches equality, an ellipse of \(\boldsymbol\theta_j^*\) is defined, although it may be clipped at zero in one or both dimensions, a complication that would decrease the effective number of degrees of freedom and which we will ignore for our purposes of estimation here.

We want to find the maximum range of \(s_j^*\) (the first dimension of \(\boldsymbol\theta_j^*\)) given \(\mathbf{F}_j\), which happens at the critical values of \(s_j^*\) given by:
\begin{align}
s_{j\pm}^{\mathrm{crit}} &= \hat{s}_j \pm \sqrt{\frac{(\Lambda_{\mathrm{threshold}} - \Lambda_j) F_{j;11}}{\det \mathbf{F}_j}},
\end{align}
and \(s_{j-}^{\mathrm{crit}}\) is clipped at zero if negative.
Note that \(F_{j;11}\) and \(\det \mathbf{F}_j\) are both non-negative because of the properties of the Hessian discussed earlier.

The final range of critical values of \(s_j^*\) are chosen as the \(z\)-sigma confidence interval for \(\hat{s}_j\).
The overall \(z\)-sigma confidence interval of \(s_{\mathrm{min}}\) is then estimated by finding the minimum and maximum critical values for all points within the spatial confidence interval; i.e., all test points where \(\Lambda_j \le \Lambda_{\mathrm{threshold}}\).


\subsection{PSL with an attenuation parameter}
The basic PSL approach from the previous section was modified to introduce a new parameter --- the photon mean free path (MFP) in any material in the scene.
The primary change is in the calculation of the response matrix element \(R_{ij}\), where the fraction of unattenuated photopeak photons \(\tau_{ij}\) is found.
To calculate \(\tau_{ij}\), the 3-D environment was represented using a regular grid of voxels that are labeled as either occupied or unoccupied according to whether they included any LiDAR points.
Restricting the analysis to only photons in the photopeak region, and assuming a single class of solid material, the attenuation model becomes dependent only on the mean free paths of the photons in any intervening material (air or solid):
\begin{align}
\tau_{ij} &= \exp\left(-r^{\mathrm{air}}_{ij} / \lambda_{\mathrm{air}} \right) \exp\left(-r^{\mathrm{solid}}_{ij} / \lambda_{\mathrm{solid}}\right),
\end{align}
where \(r^{\mathrm{air}}_{ij}\) and \(r^{\mathrm{solid}}_{ij}\) are the total lengths of air and solid material that are traversed by \(\mathbf{r}_{ij}\), and \(\lambda_{\mathrm{air}}\) and \(\lambda_{\mathrm{solid}}\) are the photon mean free paths.

\begin{figure}[t!]
\begin{center}
\begin{tikzpicture}[scale=1.2]

\definecolor{cbgray}{RGB}{127,127,127}
\definecolor{cbred}{RGB}{214,39,40}

\begin{scope}[every node/.style={minimum size={1.2cm-\pgflinewidth}, outer sep=0pt}]
    \def\occupiedx{2.5, 2.5, 3.5, 3.5, 3.5, 4.5}
    \def\occupiedy{3.5, 2.5, 2.5, 1.5, 0.5, 0.5}
    \foreach \x [count=\c,evaluate=\c as \y using {{\occupiedy}[\c-1]}]  in \occupiedx {
        \node[fill=cbgray!30] at (\x, \y) {};
    }
    \draw[step=1cm,color=black] (0,0) grid (6,4);
\end{scope}

\def\pointcloudx{3.14, 2.79, 2.58, 2.59, 3.08, 3.14, 3.90, 3.65, 3.36, 4.07, 2.88, 3.01, 3.12, 3.10, 3.09, 3.77, 3.73, 3.92, 3.50, 3.40, 2.98, 3.11, 2.97, 3.49, 3.26, 3.77, 3.71, 2.83, 2.53, 2.66, 2.71, 4.11, 4.04, 2.48, 3.21, 3.78, 3.25, 2.72, 3.66, 3.27, 3.22, 3.37, 3.31, 3.39, 3.29, 3.29, 3.17, 3.83, 2.83, 3.43, 3.40, 2.84, 2.78}
\def\pointcloudy{2.40, 3.35, 3.80, 3.70, 2.07, 2.64, 0.53, 1.20, 1.96, 0.09, 3.20, 2.96, 2.33, 2.42, 2.89, 0.22, 1.05, 0.41, 0.96, 1.41, 2.71, 2.05, 2.40, 1.11, 2.07, 0.92, 0.70, 2.74, 3.82, 3.28, 3.70, 0.04, 0.16, 3.85, 2.46, 0.60, 1.57, 3.54, 1.30, 2.36, 2.01, 2.00, 2.18, 1.27, 1.85, 1.72, 2.57, 0.11, 2.78, 1.22, 1.81, 3.07, 3.34}
\foreach \x [count=\c,evaluate=\c as \y using {{\pointcloudy}[\c-1]}]  in \pointcloudx {
    \filldraw [black] (\x, \y) circle (0.3pt);
}

\def\rayoriginx{0.2}
\def\rayoriginy{0.8}
\def\raytargetx{5.7}
\def\raytargety{3.6}

\def\rayslopeyoverx{((\raytargety - \rayoriginy) / (\raytargetx - \rayoriginx))}

\def\rayintAy{(1.0 - \rayoriginy)}
\def\rayintAx{(\rayintAy / \rayslopeyoverx)}

\def\rayintBx{(1.0 - \rayoriginx - \rayintAx)}
\def\rayintBy{(\rayintBx * \rayslopeyoverx)}

\def\rayintCx{(2.0 - \rayoriginx - \rayintAx - \rayintBx)}
\def\rayintCy{(\rayintCx * \rayslopeyoverx)}

\def\rayintDy{(2.0 - \rayoriginy - \rayintAy - \rayintBy - \rayintCy)}
\def\rayintDx{(\rayintDy / \rayslopeyoverx)}

\def\rayintEx{(3.0 - \rayoriginx - \rayintAx - \rayintBx - \rayintCx - \rayintDx)}
\def\rayintEy{(\rayintEx * \rayslopeyoverx)}

\def\rayintFx{(4.0 - \rayoriginx - \rayintAx - \rayintBx - \rayintCx - \rayintDx - \rayintEx)}
\def\rayintFy{(\rayintFx * \rayslopeyoverx)}

\def\rayintGy{(3.0 - \rayoriginy - \rayintAy - \rayintBy - \rayintCy - \rayintDy - \rayintEy - \rayintFy)}
\def\rayintGx{(\rayintGy / \rayslopeyoverx)}

\def\rayintHx{(5.0 - \rayoriginx - \rayintAx - \rayintBx - \rayintCx - \rayintDx - \rayintEx - \rayintFx - \rayintGx)}
\def\rayintHy{(\rayintHx * \rayslopeyoverx)}

\def\rayintIx{(\raytargetx - \rayoriginx - \rayintAx - \rayintBx - \rayintCx - \rayintDx - \rayintEx - \rayintFx - \rayintGx - \rayintHx)}
\def\rayintIy{(\raytargety - \rayoriginy - \rayintAy - \rayintBy - \rayintCy - \rayintDy - \rayintEy - \rayintFy - \rayintGy - \rayintHy)}


\draw[cbred, line width=1.5pt] ({\rayoriginx}, {\rayoriginy}) -- ++({\rayintAx}, {\rayintAy}) coordinate (A);
\draw[cbred, line width=1.5pt] (A) -- ++({\rayintBx}, {\rayintBy}) coordinate (B);
\draw[cbred, line width=1.5pt] (B) -- ++({\rayintCx}, {\rayintCy}) coordinate (C);
\draw[cbred, line width=1.5pt] (C) -- ++({\rayintDx}, {\rayintDy}) coordinate (D);
\draw[cbgray, line width=1.5pt] (D) -- ++({\rayintEx}, {\rayintEy}) coordinate (E);
\draw[cbgray, line width=1.5pt] (E) -- ++({\rayintFx}, {\rayintFy}) coordinate (F);
\draw[cbred, line width=1.5pt] (F) -- ++({\rayintGx}, {\rayintGy}) coordinate (G);
\draw[cbred, line width=1.5pt] (G) -- ++({\rayintHx}, {\rayintHy}) coordinate (H);
\draw[cbred, line width=1.5pt, -{Latex[length=4mm,width=2mm]}] (H) -- ({\raytargetx}, {\raytargety});

\filldraw [black] (\rayoriginx, \rayoriginy) circle (2pt);
\filldraw [black] (\raytargetx, \raytargety) circle (2pt);
\draw (A) circle (2pt);
\draw (B) circle (2pt);
\draw (C) circle (2pt);
\draw (D) circle (2pt);
\draw (E) circle (2pt);
\draw (F) circle (2pt);
\draw (G) circle (2pt);
\draw (H) circle (2pt);

\node[below=0.15cm, align=right, text width=1.5cm] at ({\rayoriginx}, {\rayoriginy}) {$\mathbf{r}_{\mathrm{start}}$};
\node[below=0.25cm, align=center, text width=1.5cm] at ({\raytargetx}, {\raytargety}) {$\mathbf{r}_{\mathrm{end}}$};

\def\xticks{0, 1, 2, 3, 4, 5, 6}
\def\xlabels{"$0$", "$\Delta_x$", "$2\Delta_x$", "$3\Delta_x$", "$4\Delta_x$", "$5\Delta_x$", "$6\Delta_x$"}
\foreach \x [count=\c,evaluate=\c as \label using {{\xlabels}[\c-1]}]  in \xticks {
    \node[below, align=center] at (\x, 0) {\footnotesize\label};
}
\node[below=0.45cm, align=center, text width=3cm] at (3, 0) {$x$ coordinate};

\def\yticks{0, 1, 2, 3, 4}
\def\ylabels{"$0$", "$\Delta_y$", "$2\Delta_y$", "$3\Delta_y$", "$4\Delta_y$"}
\foreach \y [count=\c,evaluate=\c as \label using {{\ylabels}[\c-1]}]  in \yticks {
    \node[left, align=center] at (0, \y) {\footnotesize\label};
}
\node[align=center, text width=3cm, rotate=90] at (-0.8, 2) {$y$ coordinate};

\pgfresetboundingbox
\path [use as bounding box] (-1.0, -0.8) rectangle (6.3, 4.2);
\end{tikzpicture}
\end{center}
\caption{Schematic of the ray-casting procedure, shown in two dimensions for simplicity.
The gridding is indicative of voxel edges.
Points within the shaded voxels represent point-cloud returns, thereby causing the respective containing voxels to be considered occupied.
The ray-casting algorithm starts at $\mathbf{r}_{\mathrm{start}}$ and visits each intersection point (open circles) until it reaches $\mathbf{r}_{\mathrm{end}}$. The distance traversed through occupied voxels (gray) is tallied separately from the distance traversed through unoccupied voxels (red).\label{fig:raycast}}
\end{figure}
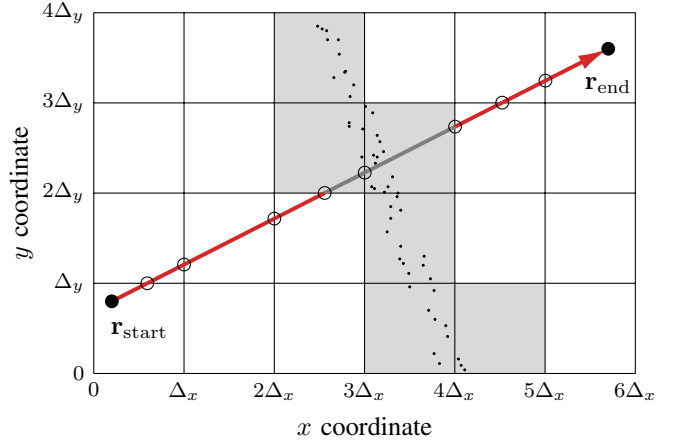

Ray casting was used to calculate the distances traversed through occupied and unoccupied voxels between any two points, and these distances were identified as \(r^{\mathrm{solid}}_{ij}\) and \(r^{\mathrm{air}}_{ij}\), respectively.
The engine used to perform the ray casting was written in Python and used the Amanatides and Woo voxel traversal algorithm~\cite{amanatides_fast_1987}.
The algorithm is described in Algorithm~\ref{alg_raycast} and illustrated in~\Fref{fig:raycast}.

\begin{algorithm}[t!]
\footnotesize
\caption{Ray-Casting Algorithm (based on~\cite{amanatides_fast_1987})}
\label{alg_raycast}
\begin{algorithmic}[1]
\Procedure{RayCast}{$\mathbf{r}_{\mathrm{start}}, \mathbf{r}_{\mathrm{end}}, \boldsymbol\Delta$}
\State $r^{\mathrm{air}} = 0$
\State $r^{\mathrm{solid}} = 0$
\State $\mathbf{k} = (\mathbf{r}_{\mathrm{end}} - \mathbf{r}_{\mathrm{start}}) / \| \mathbf{r}_{\mathrm{end}} - \mathbf{r}_{\mathrm{start}} \|$ \Comment{Calculate direction}
\State $\mathbf{r} = \mathbf{r}_{\mathrm{start}}$ \Comment{Start at origin}
\State $(i_x, i_y, i_z) = (\lfloor r_x/\Delta_x \rfloor, \lfloor r_y/\Delta_y \rfloor, \lfloor r_z/\Delta_z \rfloor)$ \Comment{Set voxel indices}
\State $d_x = 1$ if $k_x > 0$, else $0$ \Comment{Set voxel index changes}
\State $d_y = 1$ if $k_y > 0$, else $0$
\State $d_z = 1$ if $k_z > 0$, else $0$
\While{$\mathbf{r} \not= \mathbf{r}_{\mathrm{end}}$}
    \State $t_x \gets [(i_x + d_x) \Delta_x - r_x] / k_x$ \Comment{Distances to $x$/$y$/$z$ boundaries}
    \State $t_y \gets [(i_y + d_y) \Delta_y - r_y] / k_y$
    \State $t_z \gets [(i_z + d_z) \Delta_z - r_z] / k_z$
    \State $t_{\mathrm{end}} = \| \mathbf{r} - \mathbf{r}_{\mathrm{end}} \|$ \Comment{Distance to destination}
    \State $t = \min(t_x, t_y, t_z, t_{\mathrm{end}})$ \Comment{Choose shortest distance}
    \If{$\mathrm{occupied}(i_x, i_y, i_z)$} \Comment{Increment traveled distances}
        \State $r^{\mathrm{solid}} \gets r^{\mathrm{solid}} + t$
    \Else
        \State $r^{\mathrm{air}} \gets r^{\mathrm{air}} + t$
    \EndIf
    \State $\mathbf{r} \gets \mathbf{r} + t \mathbf{k}$ \Comment{Move to next point}
    \If{$t = t_x$} \Comment{Update voxel indices}
        \State $i_x \gets i_x + 2 d_x - 1$ unless $k_x = 0$
    \ElsIf{$t = t_y$}
        \State $i_y \gets i_y + 2 d_y - 1$ unless $k_y = 0$
    \ElsIf{$t = t_z$}
        \State $i_z \gets i_z + 2 d_z - 1$ unless $k_z = 0$
    \EndIf
\EndWhile
\EndProcedure
\end{algorithmic}
\end{algorithm}

With \(r_{ij}^{\mathrm{solid}}\) calculated, it remains to find the value of  \(\lambda_{\mathrm{solid}}\).
The value of \(\lambda_{\mathrm{air}}\) can be obtained from tables using the photopeak energy and assuming dry air at standard temperature and pressure; \(\lambda_{\mathrm{air}}=\)107.8\,m and 83.4\,m were used for the analysis of 662\,keV and 356\,keV photopeak events, respectively~\cite{mcconn_jr_compendium_2011, xcom}.
The approach taken was to treat \(\lambda_{\mathrm{solid}}\) as another parameter of PSL to optimize over, in addition to \(s\), \(b\), and position (\(x\), \(y\), \(z\)).
In this proof-of-concept, the optimization was done in a brute-force manner --- a grid of \(\lambda_{\mathrm{solid}}\) values was chosen and PSL was solved over a four-dimensional grid of (\(x\), \(y\), \(z\), \(\lambda_{\mathrm{solid}}\)).
The test point index \(j\) was expanded to denote both spatial position and \(\lambda_{\mathrm{solid}}\) value, and then the PSL solution with the smallest \(-\log \hat{L}_{\mathrm{min}}\) was chosen as the best overall point-source model.
Because we now effectively have six degrees of freedom instead of five, a \(\chi^2_6\) distribution was used when calculating the confidence intervals (i.e., using the LRT threshold in~\Fref{eq:lambda_threshold}).
When calculating confidence intervals, all PSL solutions in the four-dimensional space that were below the likelihood threshold were used.
Analogous to spatial confidence intervals, a confidence interval for \(\lambda_{\mathrm{solid}}\) can likewise be generated.


\subsection{Validation of the approach using a toy model}\label{sec:toy_model}
\begin{figure}[t!]
\begin{center}
\includegraphics[width=0.99\columnwidth]{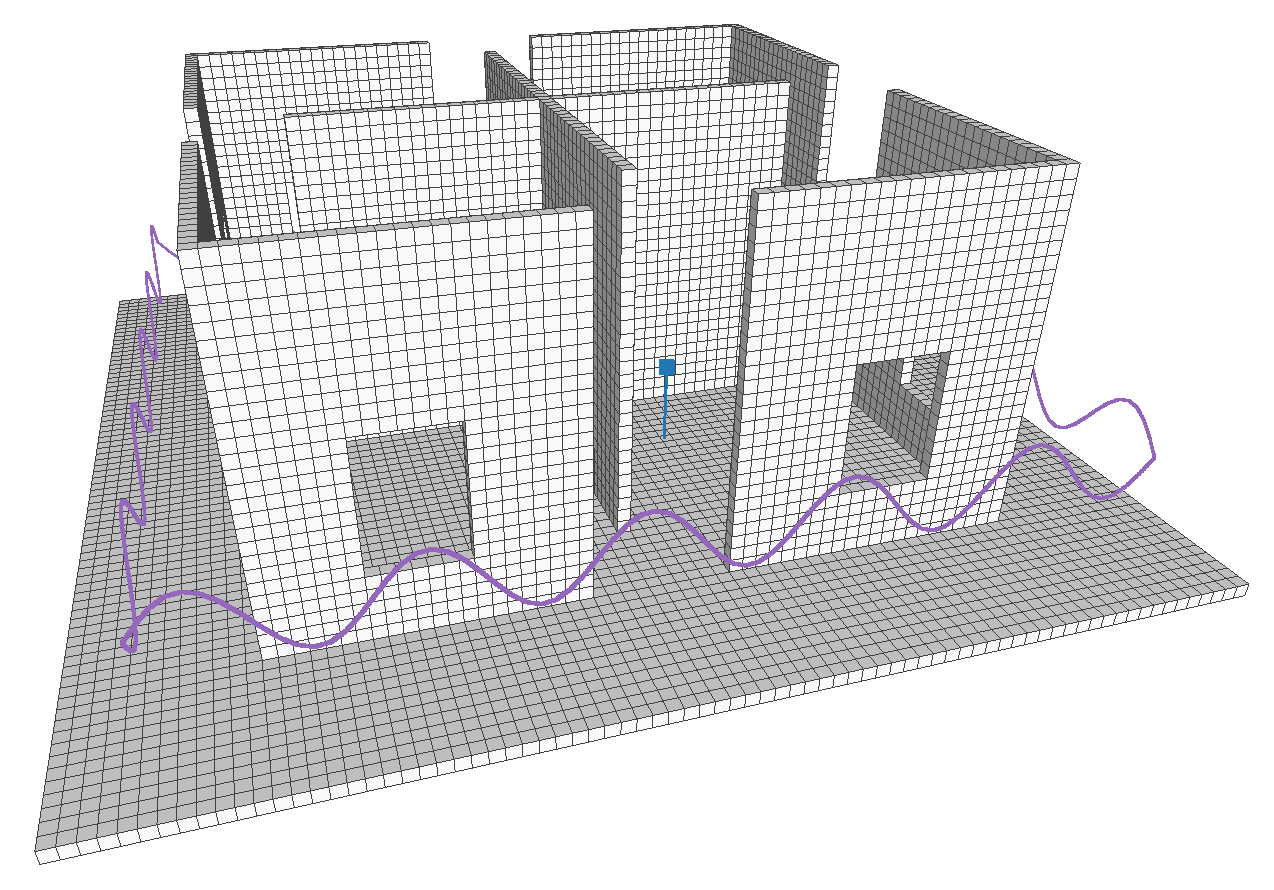}
\end{center}
\caption{A three-dimensional rendering of the toy model used, showing the path of the detector (purple) and the position of the source (blue dot, with vertical line to indicate horizontal location).\label{fig:toy_model_screenshot}}
\end{figure}

A toy model was developed to simulate the effect of a point source inside a simple structure with external walls, internal walls, windows, and doors.
The building was constructed using a regular 3-D voxel space with a voxel size of 15\,cm.
It consisted of an 8\(\times\)8-m external wall that was one voxel thick, with some portions of the wall removed to represent windows and doors.
Internal walls were added in the form of one-voxel thick structures.
The material making up the building was assumed to have an MFP of 0.30\,m for 662\,keV photons, which is roughly the value for low density wood.
A hypothetical detector with a spherically symmetric response function (\(A =\)~10\,cm\(^2\)) was moved along a 10\(\times\)10-m square surrounding the structure over the course of 60~seconds while traveling at a constant speed, and data were recorded at a rate of 5\,Hz.
The height of the detector varied as a sinusoid of amplitude 40\,cm to simulate the motion of walking and add variation in the vertical dimension.
A 500-\textmu Ci \isot{Cs}{137} point source was placed inside the structure 1~meter above the floor at the same mean height as the detector.
The resulting structure, detector path, and source location is shown in~\Fref{fig:toy_model_screenshot}.

\begin{figure}[t!]
\begin{center}
\includegraphics[width=0.95\columnwidth]{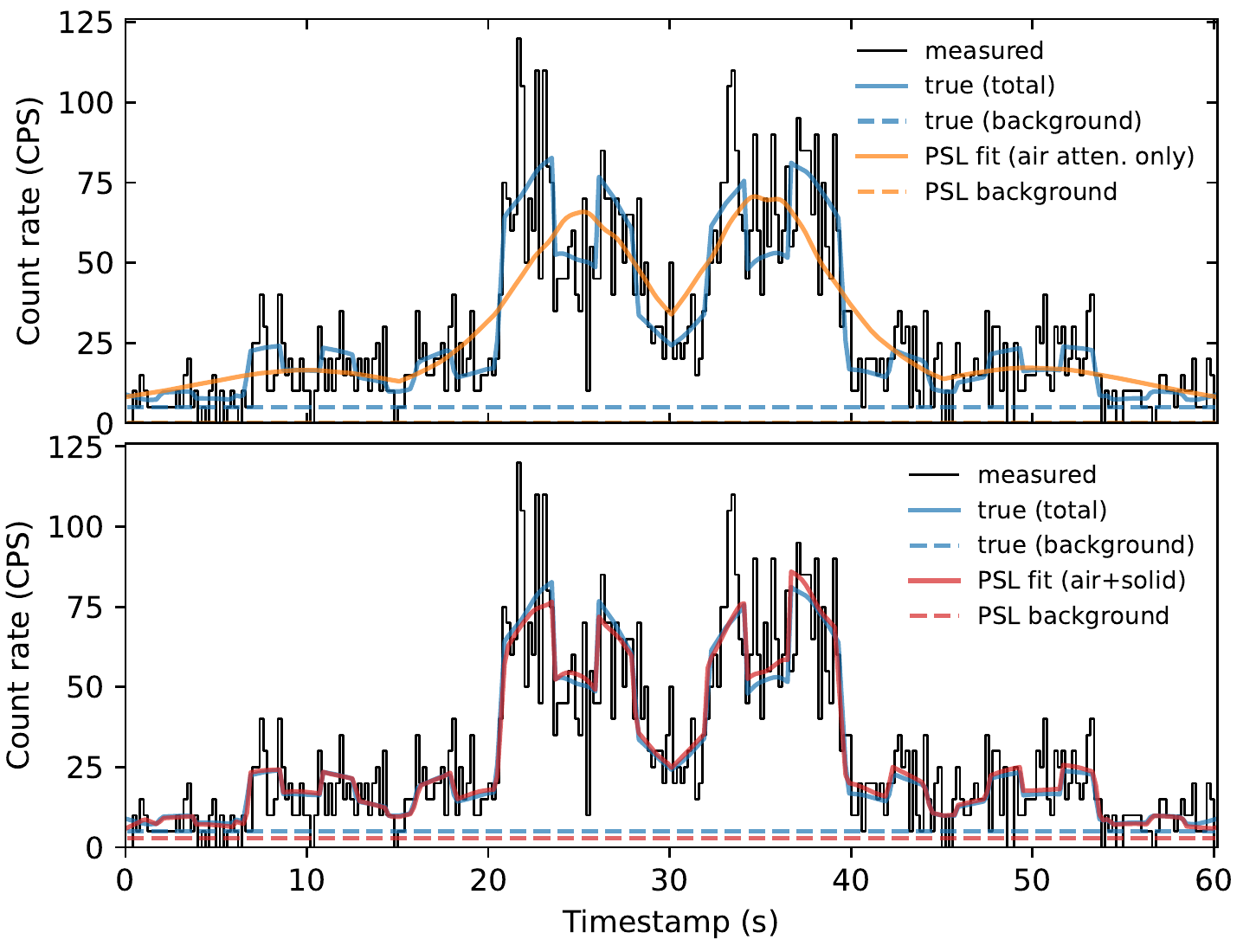}\\
\includegraphics[width=0.90\columnwidth]{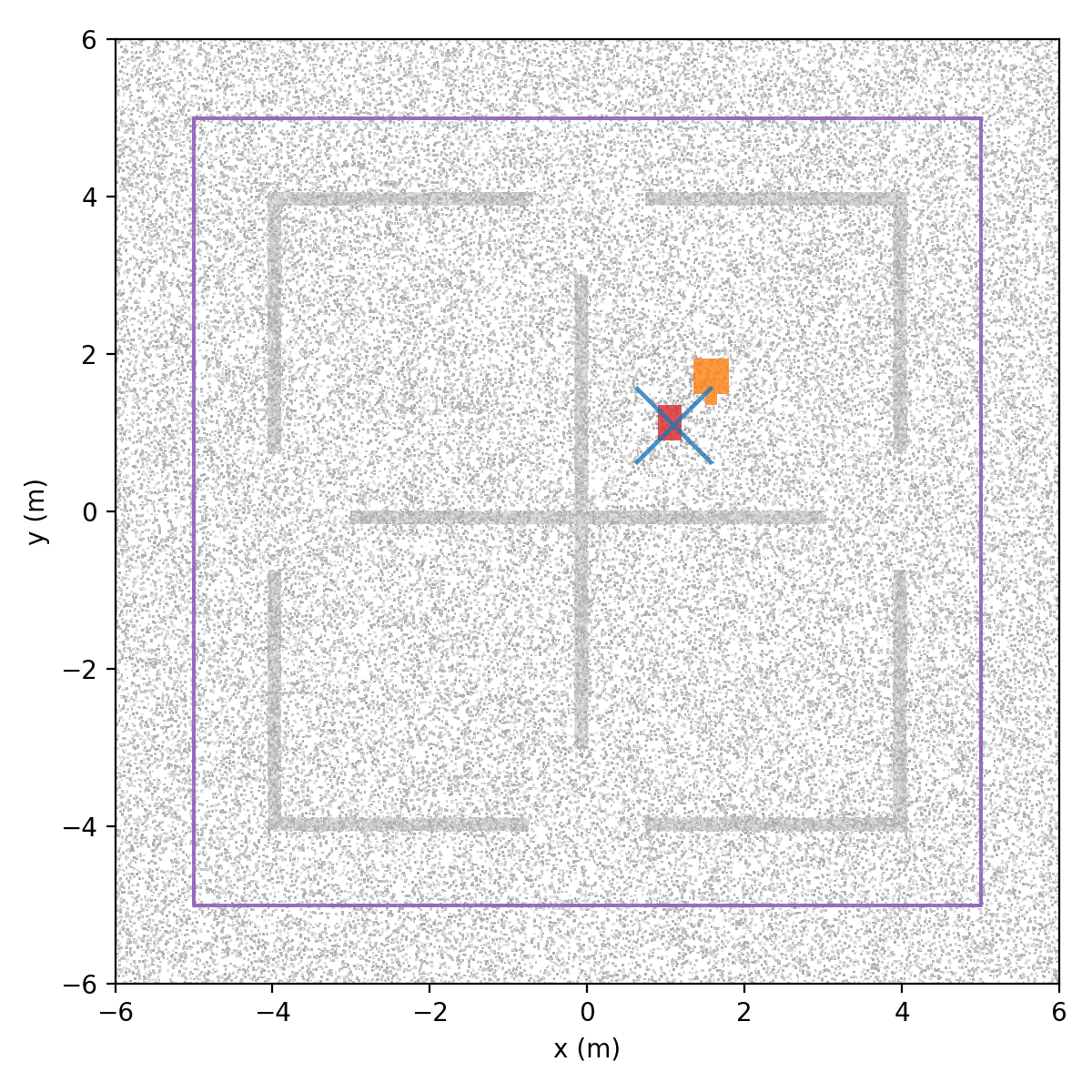}\\
\includegraphics[width=0.90\columnwidth]{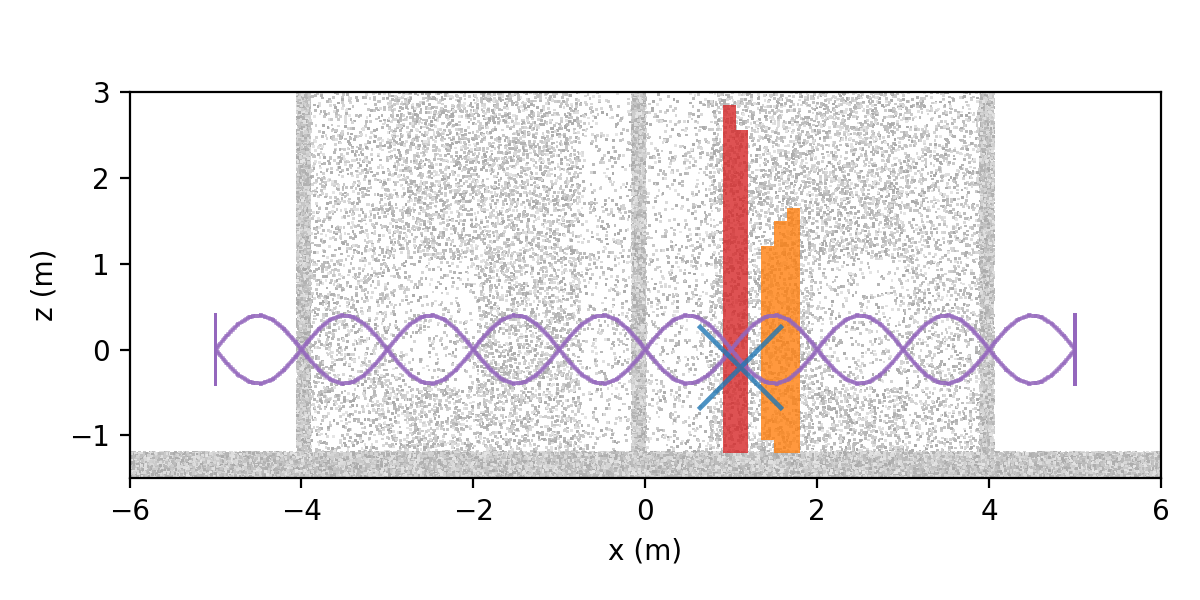}
\end{center}
\caption{The simulated counts in the detector for the toy model, showing the best fit when only air attenuation is included (top) and when a solid attenuation parameter is fit (second from top).
The resulting spatial confidence intervals are shown in the bottom two figures, where the 2-sigma confidence intervals are shown as red and orange volumes, for the with-solid and air-only attenuation models, respectively.
The simulated trajectory of the system is shown in purple and the true source location is shown as a blue cross.
The point cloud in these plots is a simulated point cloud based on the occupied voxels of the toy model.\label{fig:toy_model_counts}}
\end{figure}

PSL was performed with and without fitting the attenuation from the structure, using all voxel centers inside the structure as test points.
To optimize over MFP, PSL was solved at 100~values spaced logarithmically from 5\,cm to 100\,m.
The shape of the best fit negative log likelihood as a function of MFP was concave, and thus there was only one local optimum for \(\lambda_{\mathrm{solid}}\), which was also the global optimum, which is 0.37\,m.
Note, however, that the negative log likelihood function is not in general guaranteed to be concave along this dimension, nor in any of the three spatial dimensions.
The algorithm is able to provide a 2\textsigma~confidence interval for the mean free path in the same way as it does for the spatial dimensions by noting the range of MFP values for which any test points exist with \(\Lambda\) values below \(\Lambda_{\mathrm{threshold}}\); for the toy model, this interval is 0.25--0.58\,m.
The best fit source activity was 525\,\textmu Ci, with a 2\textsigma~confidence interval of 440--760\,\textmu Ci, which is in agreement with the simulated activity.
In addition, the 2\textsigma~spatial confidence interval enclosed the true source position.
In contrast, the PSL solution that ignored attenuation from occupied voxels returned a 2\textsigma~spatial confidence interval whose nearest boundary was approximately 40\,cm from the true position and gave a source activity confidence interval of 270--380\,\textmu Ci, which excludes the true activity.
\Fref{fig:toy_model_counts} shows the simulated count rates and the 2-sigma bounds of the most probable source locations in the point cloud for the best PSL solutions when attenuation is optimized versus set to \(\lambda_{\mathrm{air}}\).

\section{Results for experimental data}\label{sec:results}
In order to test the approach, experimental data were taken on 17~August 2020 with sources in a realistic scenario.
The scenario consisted of several stacked 20-foot shipping containers in a field at the Richmond Field Station (RFS) in Richmond, CA\@.
Two sources were placed in the scene: a \isot{Cs}{137} source with an activity of 1848\,\textmu Ci was placed on the outside of a container in a narrow gap between the containers, and a 638-\textmu Ci \isot{Ba}{133} source placed inside on the floor of an empty container in a second grouping of containers.
The sources were approximately 30\,m away from each other, and the path of the system allowed two subsets of the data to be chosen that effectively separated the source measurements into two independent measurements.

\begin{figure}[t!]
\begin{center}
\begin{tikzpicture}
    \node[anchor=south west, inner sep=0] (image) at (0,0) {\includegraphics[width=0.9\columnwidth, trim={0 2cm 0 8cm}, clip]{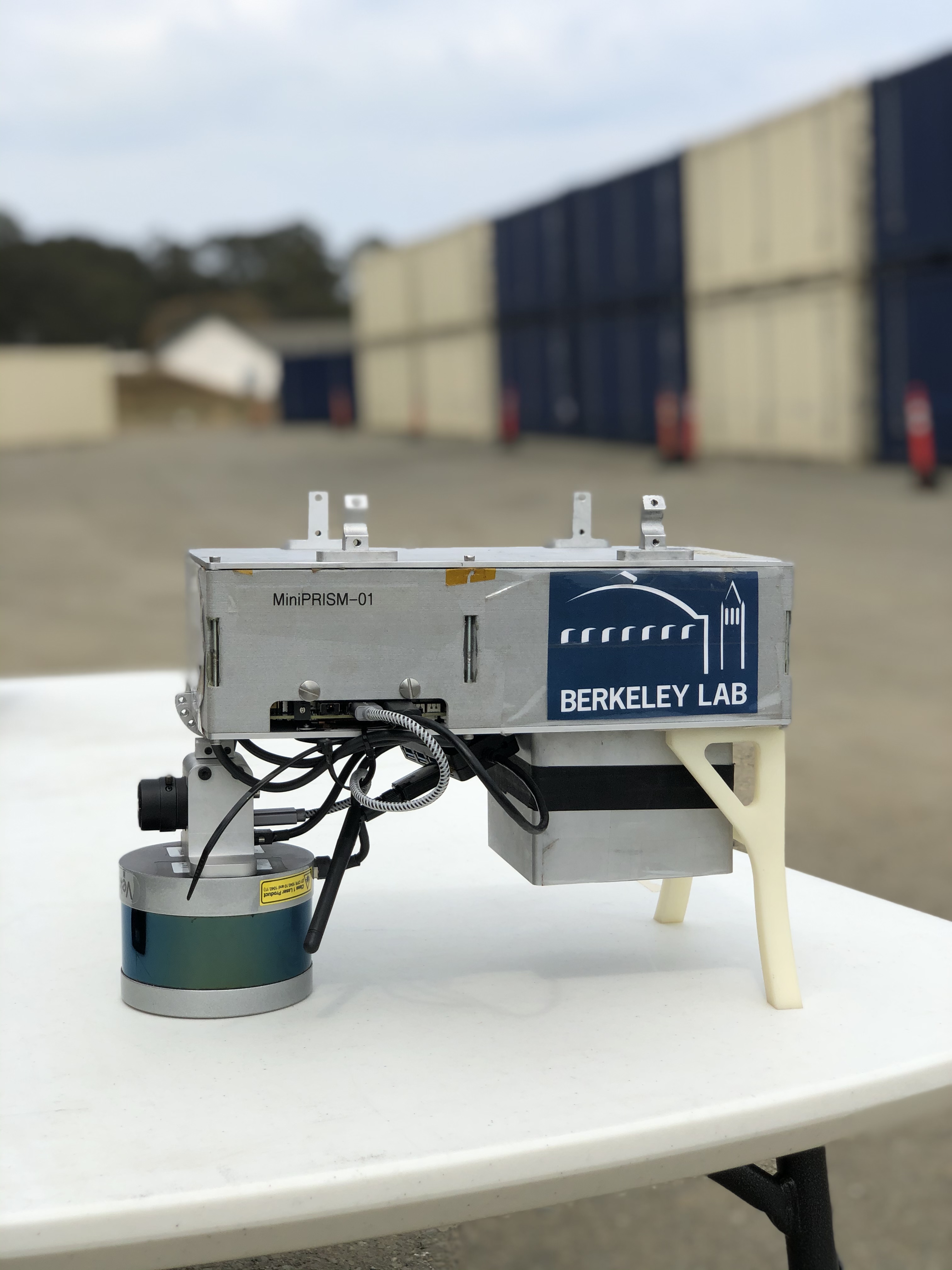}};
    \begin{scope}[x={(image.south east)}, y={(image.north west)}]
        \node[red] (camera) at (0.08, 0.55) {\small Camera};
        \draw[red, -latex] (camera) -- ++(0.07, -0.10);
        \node[red] (lidar) at (0.1, 0.1) {\small LiDAR};
        \draw[red, -latex] (lidar) -- ++(0.12, 0.12);
        \node[red] (imu) at (0.5, 0.2) {\small IMU};
        \draw[red, -latex] (imu) -- ++(-0.23, 0.19);
        \node[red] (minip) at (0.8, 0.15) {\small CZT array};
        \draw[red, -latex] (minip) -- ++(-0.15, 0.25);
        \fill[white, opacity=0.5] (0.01, 0.82) rectangle (0.41, 0.88);
        \node[red] (battery) at (0.21, 0.85) {\small Battery and electronics};
        \draw[red, -latex] (battery) -- ++(0.07, -0.20);
        \fill[white, opacity=0.5] (0.55, 0.87) rectangle (0.95, 0.93);
        \node[red] (mount) at (0.75, 0.9) {\small UAS mounting points};
        \draw[red, -latex] (mount) -- ++(-0.06, -0.13);
    \end{scope}
\end{tikzpicture}
\end{center}
\caption{The MiniPRISM detector system before being mounted on the UAS platform.\label{fig:miniprism}}
\end{figure}
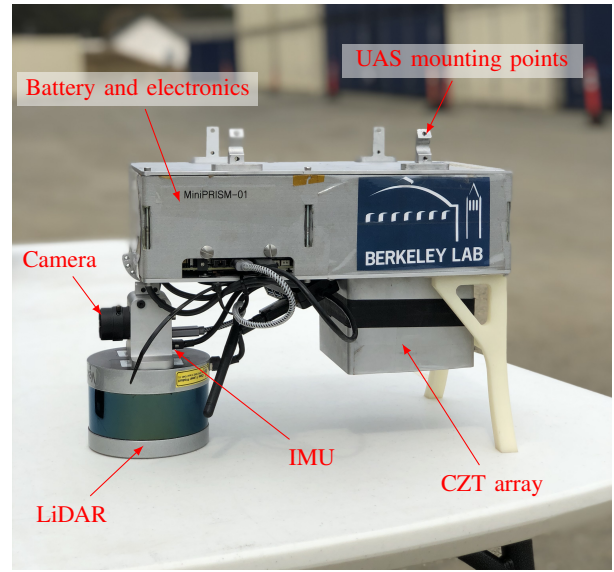

The data were collected with the MiniPRISM system, which consists of up to 64~coplanar grid CdZnTe detectors, each of which is a 1\,cm\(^3\)~cube~\cite{pavlovsky_miniprism_2019}.
MiniPRISM's detectors are arranged into a partially filled 6\(\times\)6\(\times\)4 grid of potential detector locations, with the detectors in an optimized arrangement for detection sensitivity and active coded mask imaging performance.
The detector system is integrated with a package of contextual sensors, including a LiDAR, Inertial Measurement Unit (IMU), and camera (\Fref{fig:miniprism}) and mounted on an small Unmanned Aerial System (sUAS) platform.
Although the data are individually read out by detector, for this analysis the detector events were collected together as if the system were a single, monolithic detector to maintain simplicity and keep computation times manageable while demonstrating efficacy in modeling attenuation.
A measurement-validated simulated response function in all \(4 \pi\)~steradians was used when performing PSL~\cite{hellfeld_free-moving_2021}, and the simulations used the actual configuration of 58~detectors used in the measurement.
Only the responses for the 54~detectors that had adequate performance at the time were summed together and used in the analysis.
The on-board LiDAR and IMU were used to perform Simultaneous Localization and Mapping (SLAM) using Google Cartographer~\cite{google_cartog_2016}.
SLAM solves for a fixed frame and outputs the path and orientation of the system within that frame, as well as LiDAR points mapped into that frame.
SLAM is performed in real-time by the on-board computer, and PSL without attenuation and other data products are served to the user in real-time.

\subsection{Cs-137 in gap between shipping containers}
\begin{figure*}[t!]
\begin{center}
  \begin{tikzpicture}
      \node[anchor=south west, inner sep=0] (image) at (0,0) {\includegraphics[width=0.41\textwidth, trim={3cm 0 3cm 0}, clip]{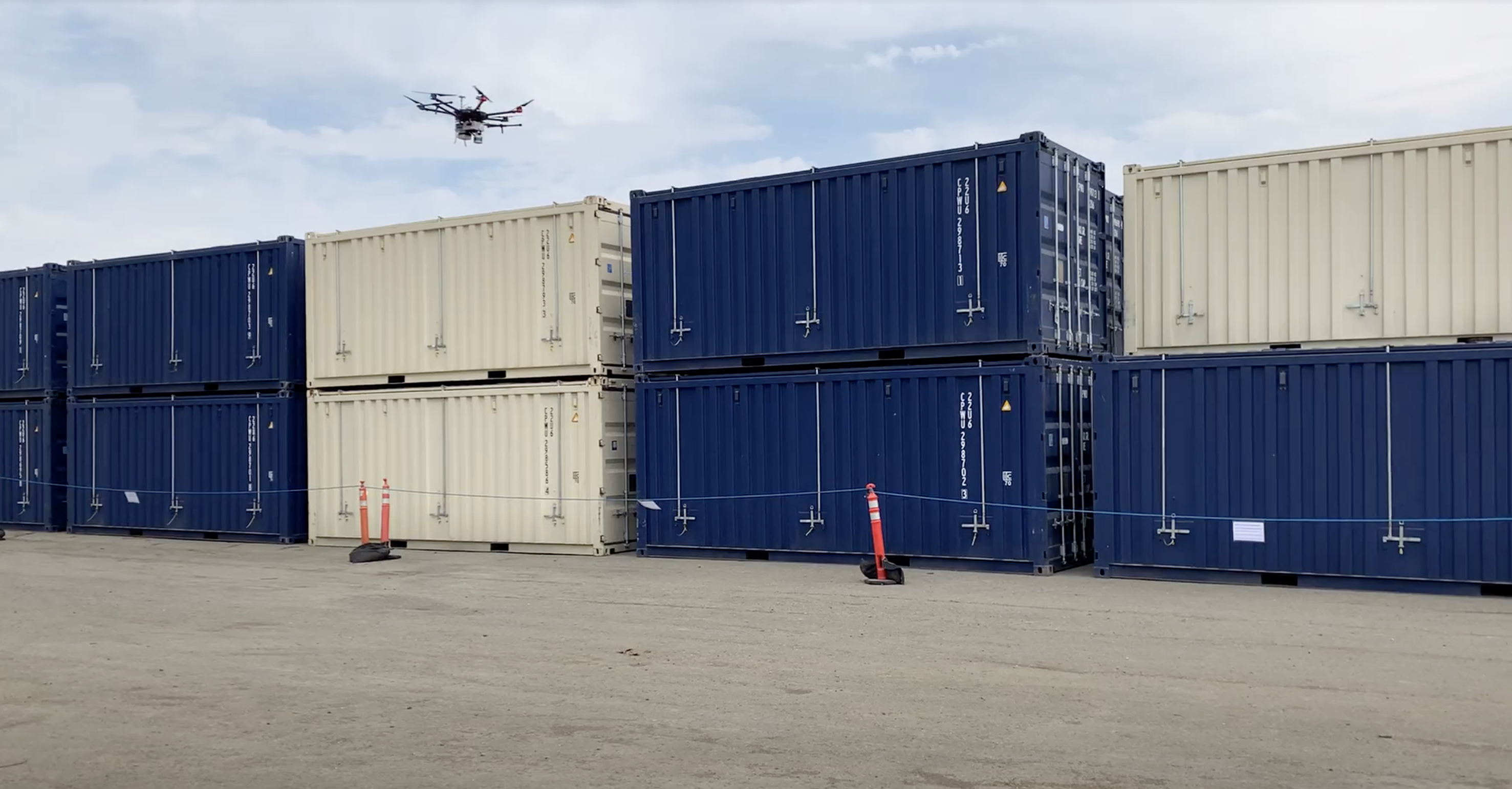}};
      \begin{scope}[x={(image.south east)}, y={(image.north west)}]
        \node[red] (uas) at (0.48, 0.90) {\small sUAS};
        \draw[red, -latex] (uas) -- ++(-0.14, -0.05);
      \end{scope}
  \end{tikzpicture}
  \hspace{4pt}
  \begin{tikzpicture}
      \node[anchor=south west, inner sep=0] (image) at (0,0) {\includegraphics[width=0.37\textwidth, trim={11cm 0 0 3cm}, clip]{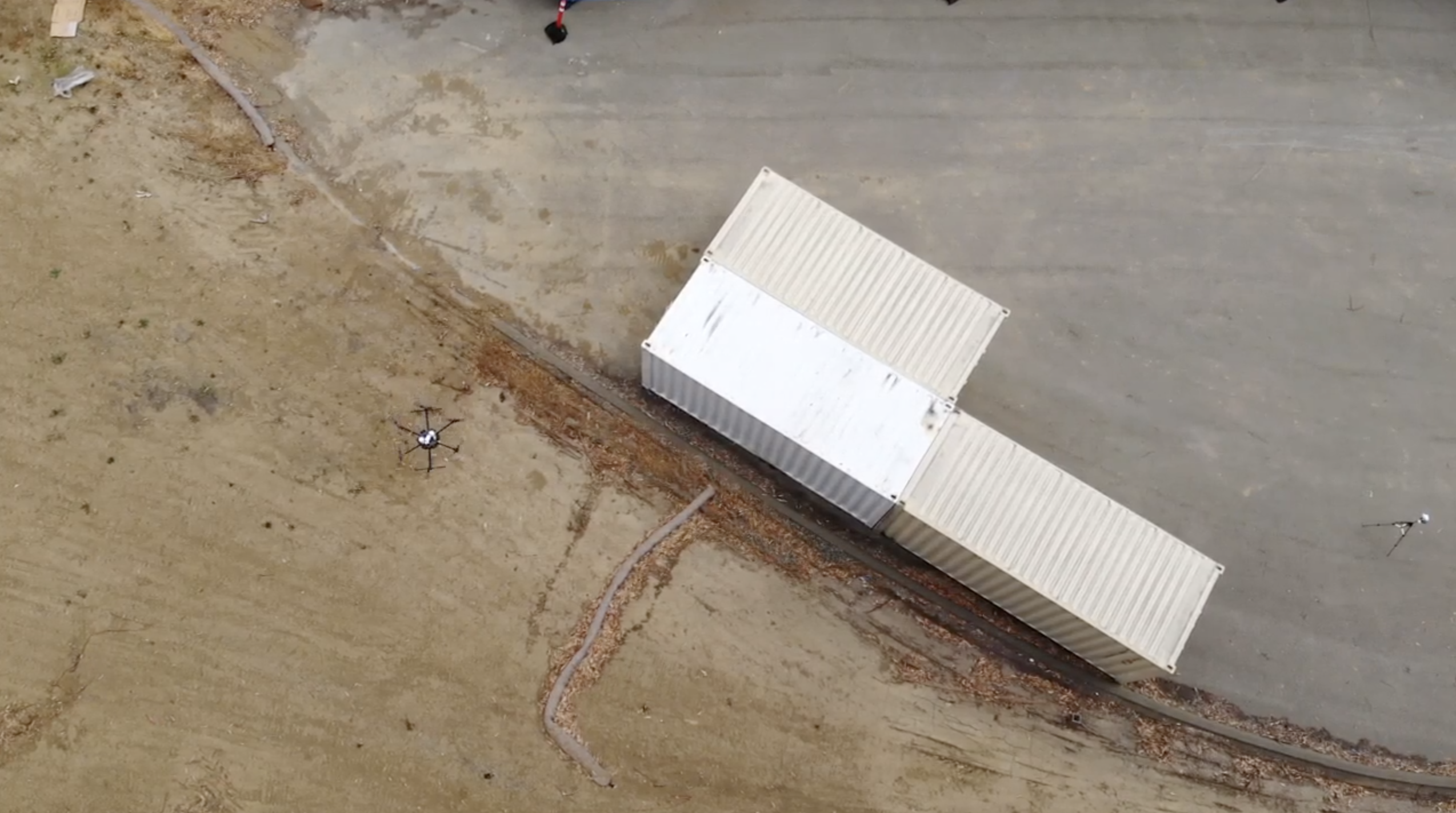}};
      \begin{scope}[x={(image.south east)}, y={(image.north west)}]
        \fill[white, opacity=0.5] (0.025, 0.25) rectangle (0.17, 0.35);
        \node[red] (uas) at (0.10, 0.30) {\small sUAS};
        \draw[red, -latex] (uas) -- ++(-0.02, +0.15);
      \end{scope}
  \end{tikzpicture}\\
  \vspace{2pt}
  \includegraphics[width=0.41\textwidth, trim={5cm 1cm 5cm 1cm}, clip]{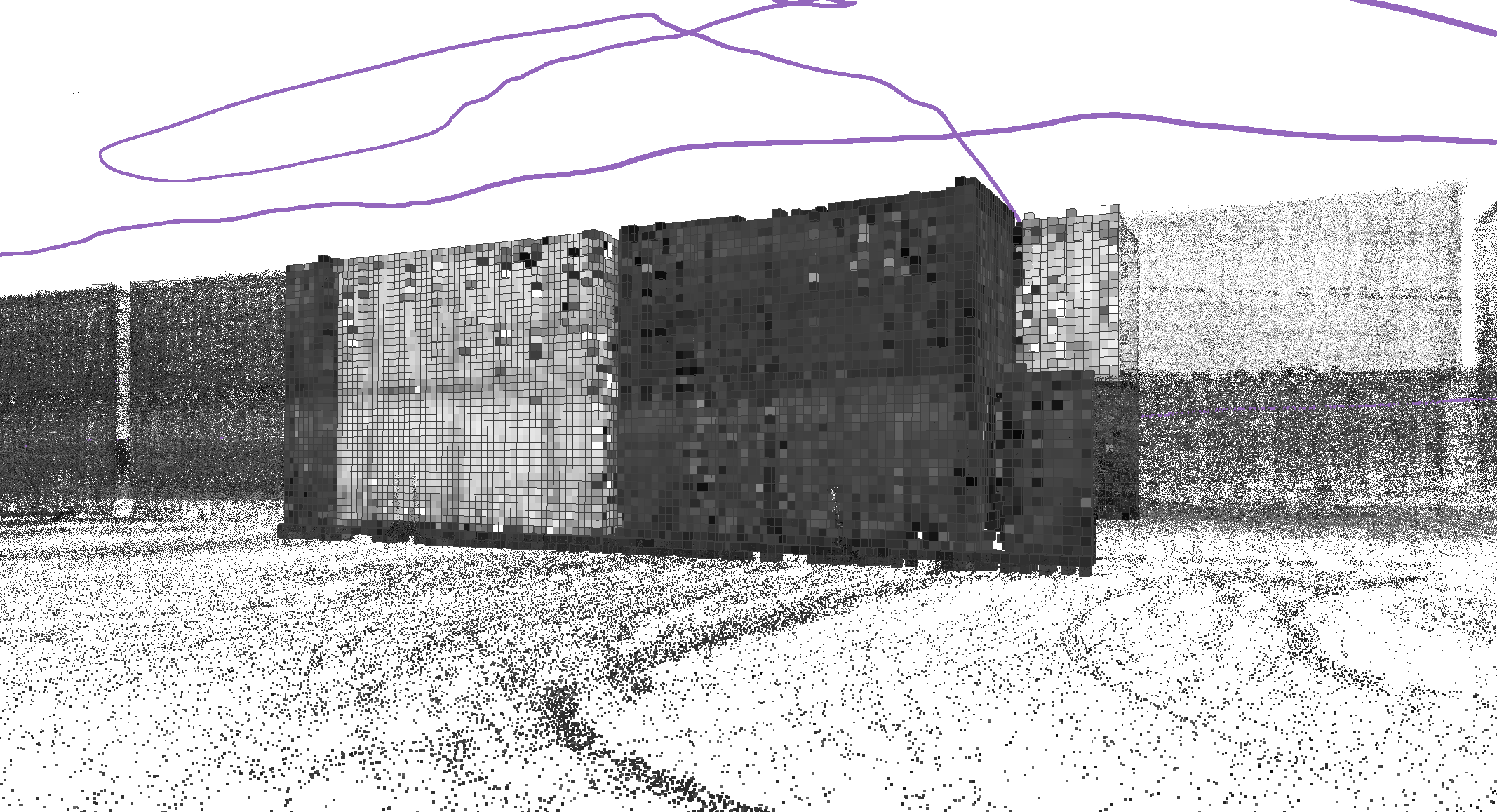}
  \hspace{4pt}
  \includegraphics[width=0.37\textwidth, trim={6cm 0cm 4cm 2cm}, clip]{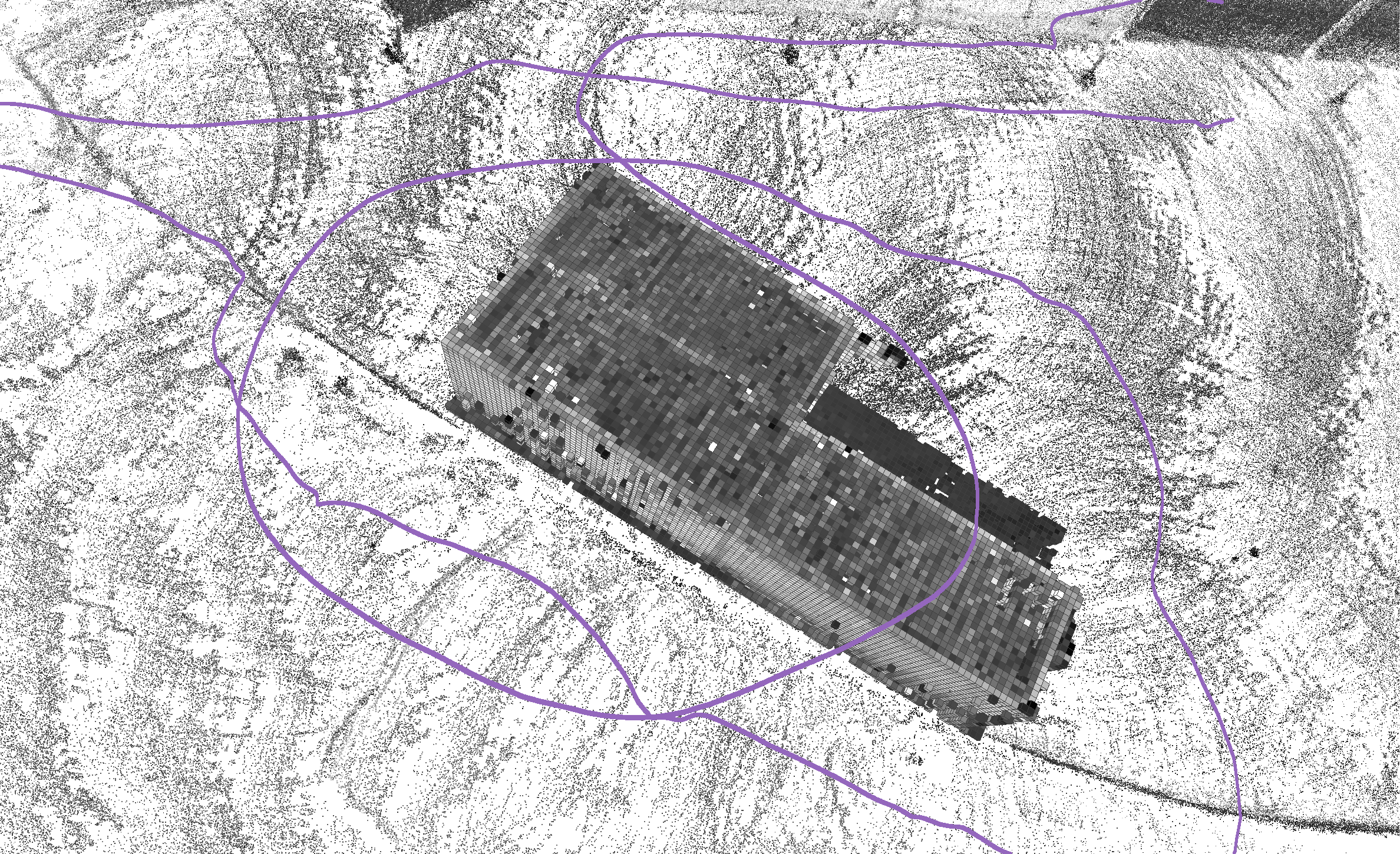}
\end{center}
\caption{The occupied voxel models made from the LiDAR point clouds.
The \isot{Cs}{137} source scenario (top left), voxelized at 13.5\,cm resolution (bottom left).
The \isot{Ba}{133} source scenario (top right), voxelized at 12\,cm resolution (bottom right).
The purple path is the path of the detector system, and the grayscale points are the LiDAR point cloud colorized by near-infrared reflectivity.
In both of the top images, the sUAS carrying MiniPRISM is indicated.\label{fig:voxel_models}}
\end{figure*}
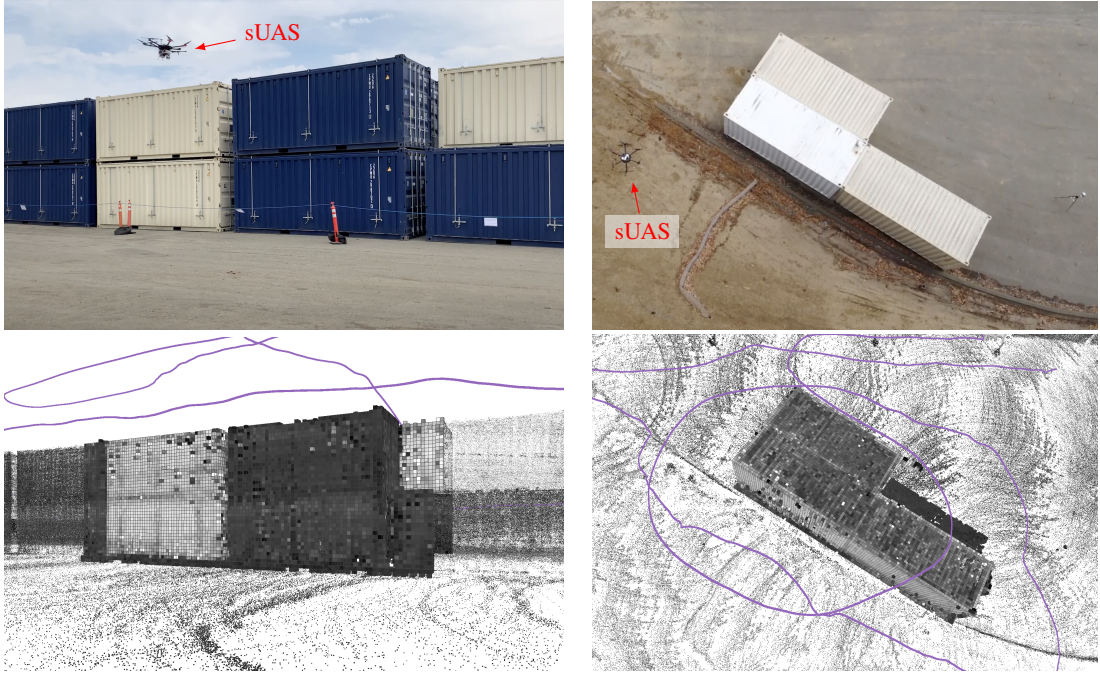

To search for the \isot{Cs}{137} source, the sUAS was flown along both sides of the container stacks at an average standoff of 4\,m from the sides of the containers and at an altitude of 3\,m above ground level (AGL), and then flown at 8\,m AGL over the center of the stacks (and about 3\,m above the top of the stack) where the source was located, for a total duration of about 220\,s.
Data were binned at 5\,Hz for analysis, and the spectral region from 620--700\,keV was selected to capture the 662\,keV photopeak.\footnote{Although the detectors are CdZnTe and can in principle achieve an energy resolution of 2\% FWHM~\cite{luke_factors_2004}, the individual channel nonlinearities and gain drift were difficult to correct to the level of the ideal resolution. In addition, any non-photopeak events can be fit as background by PSL.}
A voxel grid with 13.5-cm pitch was rotated and translated by hand to align it with the container stack and exploit its rectilinear shape, and any voxels containing points from the LiDAR point cloud were marked as occupied.
Those voxels within individual containers were not considered occupied.
The resulting occupied voxel model is shown in~\Fref{fig:voxel_models}.

\begin{figure}[t!]
\begin{center}
  \includegraphics[width=0.99\columnwidth]{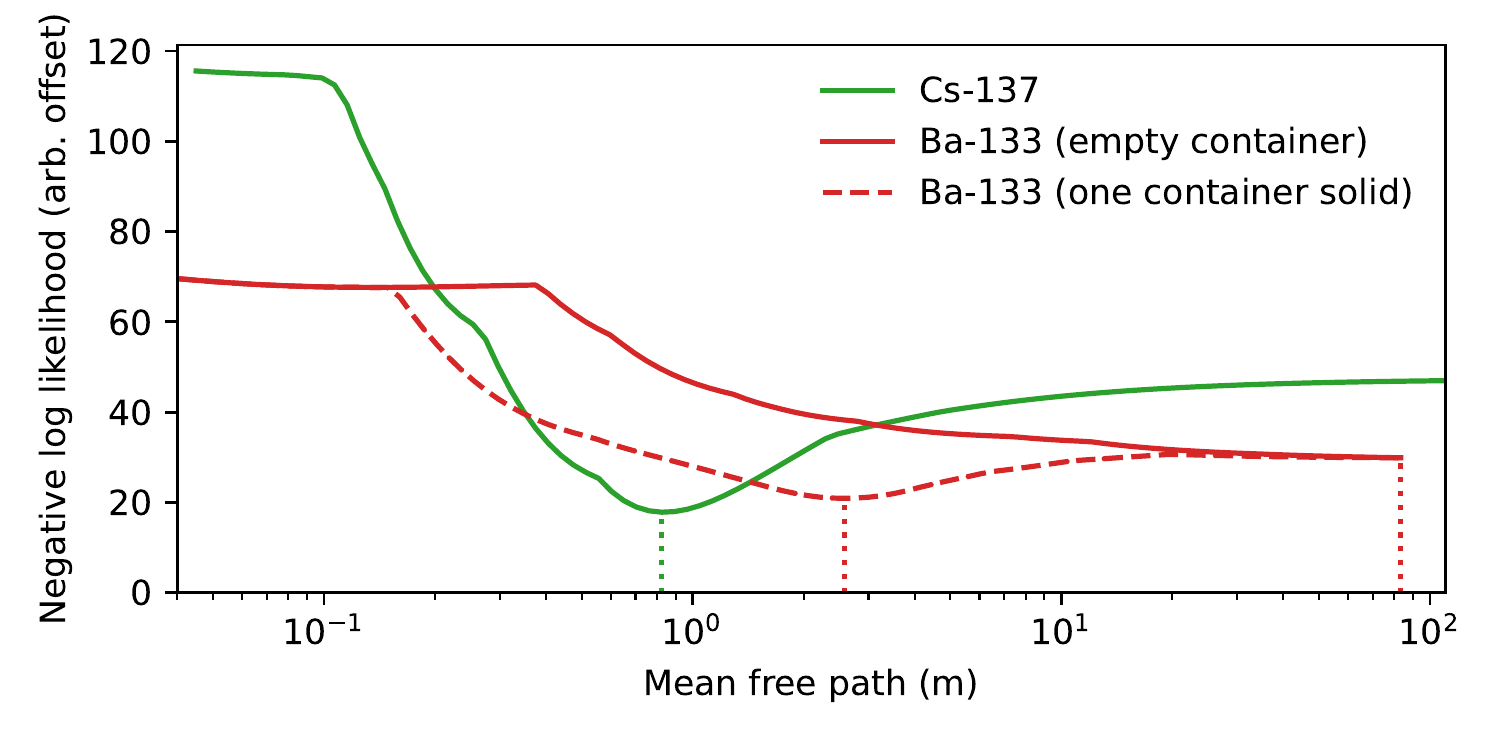}
\end{center}
\caption{The best negative log likelihood values as a function of \(\lambda_{\mathrm{solid}}\) for the experimental data.\label{fig:nll_vs_mfp}}
\end{figure}

\begin{figure}[t!]
\begin{center}
  \includegraphics[width=0.99\columnwidth]{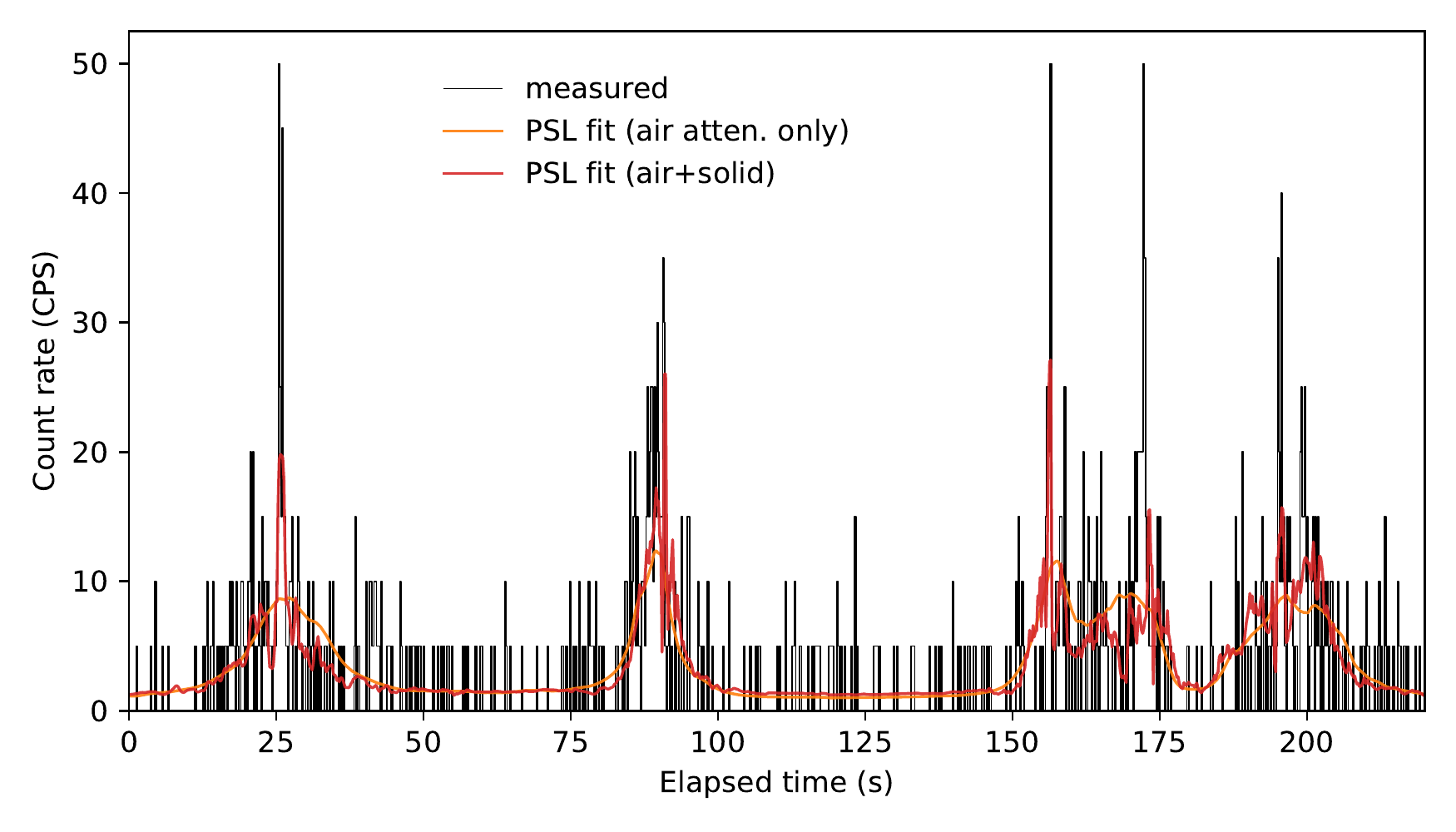}\\
  \includegraphics[width=0.99\columnwidth]{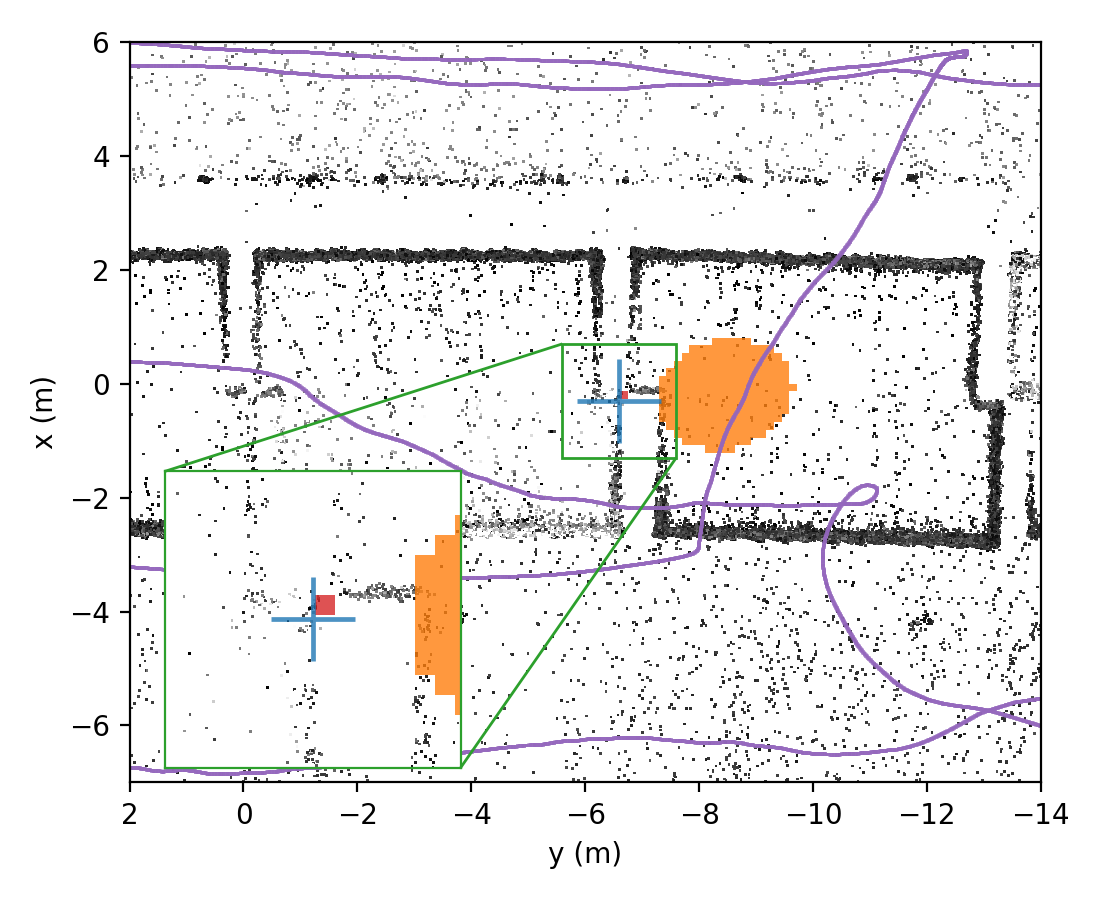}\\
  \includegraphics[width=0.99\columnwidth]{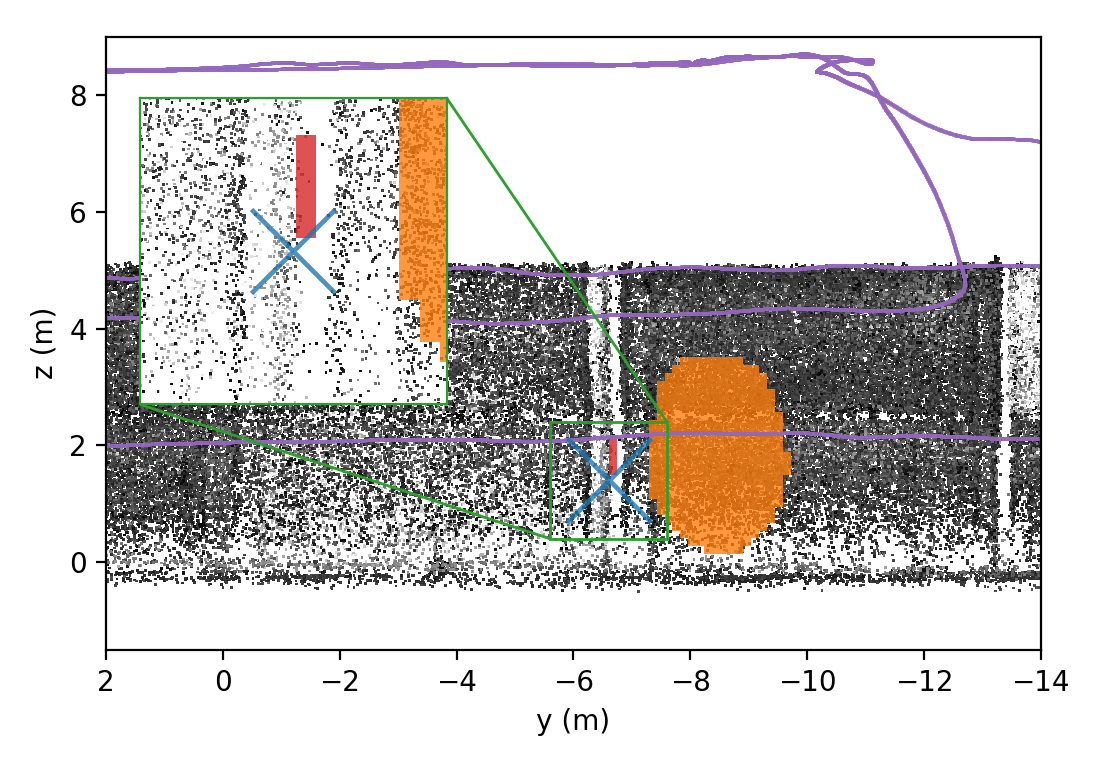}
\end{center}
\caption{Results for \isot{Cs}{137} showing the best fit model when no attenuation is assumed for occupied voxels (orange) and when the optimal MFP is found for occupied voxels (red).
The top plot shows the count rates, while the bottom two plots show the 2\textsigma~spatial confidence intervals for top-down and side views in the same colors as the count rate plot.
The ground truth is shown with a blue cross.
Insets are given to zoom in on the 2\(\times\)2\,m region around the ground truth.
Note that the red region is small, taking up only five voxels, and localizes the source to within the gap between the containers.\label{fig:cs137_results}}
\end{figure}

PSL was performed using the centers of all voxels within the grid as test points and optimizing \(\lambda_{\mathrm{solid}}\) over 100~values logarithmically spaced from 0.045\,m to 107.8\,m (\(\lambda_{\mathrm{air}}\) for 662\,keV).
The ray-casting calculation took 60~minutes on a four-core Intel~i7 CPU, and each MFP optimization step took 90\,s, for a total of 210~minutes for the entire optimization.
As in the toy model, the curve of negative log likelihood versus MFP was found to be concave (the blue line in~\Fref{fig:nll_vs_mfp}).
The optimal MFP obtained was 0.82\,m, with a 2\textsigma~confidence interval of 0.60--1.32\,m.
The most likely source activity was 1440\,\textmu Ci, with a 2\textsigma~confidence interval of 960--2070\,\textmu Ci, which includes the true activity at the time the measurement was made (1848\,\textmu Ci).
Without attenuation from occupied voxels, the confidence interval is 380--730\,\textmu Ci, which excludes the true activity.
The 2\textsigma~spatial confidence interval was limited to only five 13.5\,cm voxels, all falling inside the gap between containers where the source was actually located.
The source ground truth position fell just outside of these five voxels by about 10\,cm, which is on the order of the voxelization and fidelity of the ground truth itself.
These results are shown in \Fref{fig:cs137_results}.

The best fit value of \(\lambda_{\mathrm{solid}}\) seems at first to be far too large --- the containers are made of steel, so na\"{i}vely the MFP should be approximately the tabulated value 0.0176\,m for 662\,keV~\cite{mcconn_jr_compendium_2011, xcom}.
This apparent discrepancy will be addressed in~\Fref{sec:discussion}.

\subsection{Ba-133 in an empty container}
For the \isot{Ba}{133} source, since the second stack was only a single layer of three containers, the UAS flew around the stack once at approximately 2\,m above ground level and roughly 3\,m standoff from the sides of the containers, and a second time at 5\,m AGL, which was 2\,m above the top of the containers, for a total duration of about 145\,s.
This grouping of containers consisted of three containers only in a single layer (\Fref{fig:voxel_models}).
Data were binned at 5\,Hz for analysis, and the spectral region from 315--385\,keV was selected to capture the 356\,keV photopeak.
Because the spatial region of interest was smaller than in the \isot{Cs}{137} case, a finer voxel pitch of 12\,cm was possible.
Once again the voxel grid was aligned to the containers to exploit their rectangular shape.

For PSL, \(\lambda_{\mathrm{solid}}\) was swept over 100~values logarithmically spaced from 0.04\,m to \(\lambda_{\mathrm{air}}\)=83.4\,m.
The ray-casting calculation took 30~minutes on the same system as the \isot{Cs}{137} analysis, and each MFP optimization step took 40~seconds, for a total of 100~minutes for the entire optimization.
Unlike the toy model and the \isot{Cs}{137} data, the curve of negative log likelihood versus MFP was not concave.
In addition, the negative log likelihood generally decreased with increasing MFP, leading to an optimal MFP of 83.4\,m, or essentially no attenuation from occupied voxels.

\begin{figure}[t!]
\begin{center}
  \includegraphics[width=0.99\columnwidth]{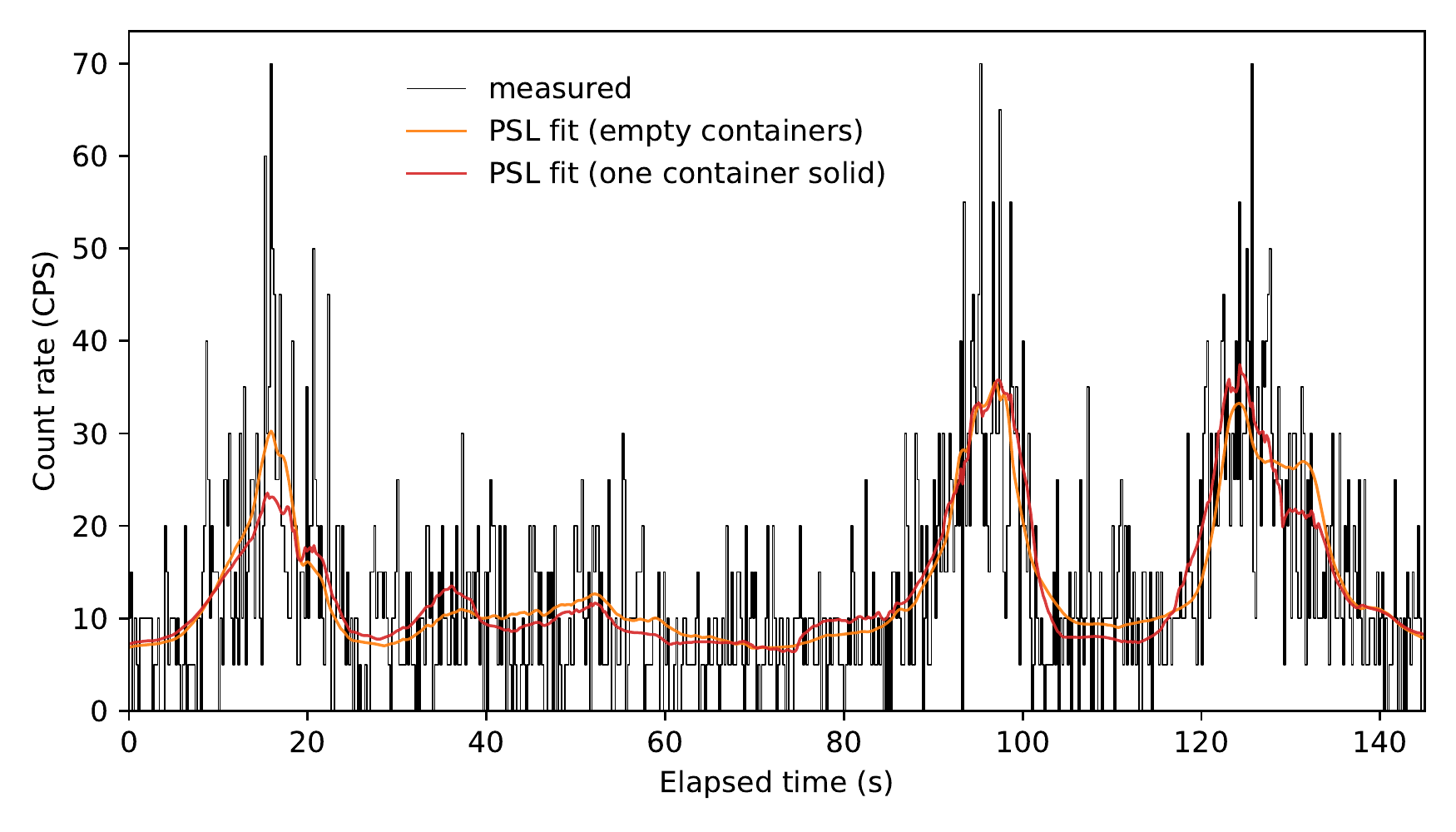}\\
  \includegraphics[width=0.99\columnwidth]{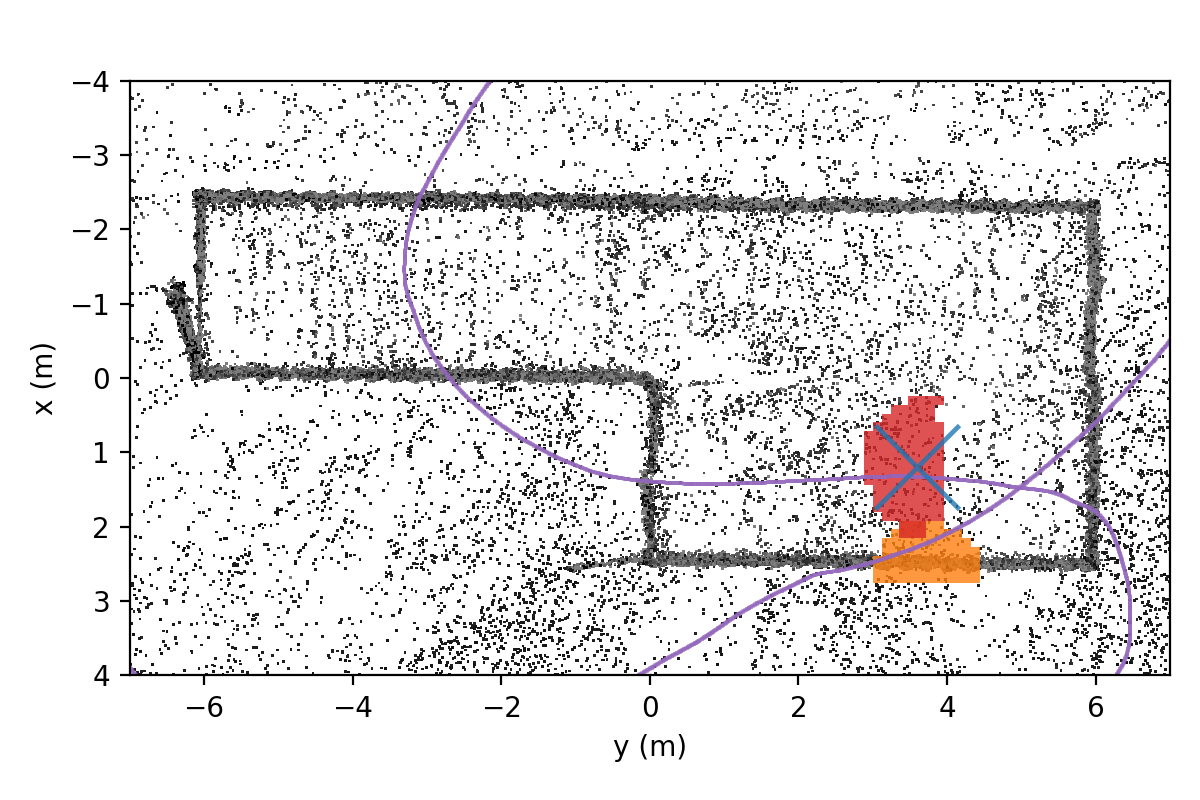}\\
\includegraphics[width=0.99\columnwidth]{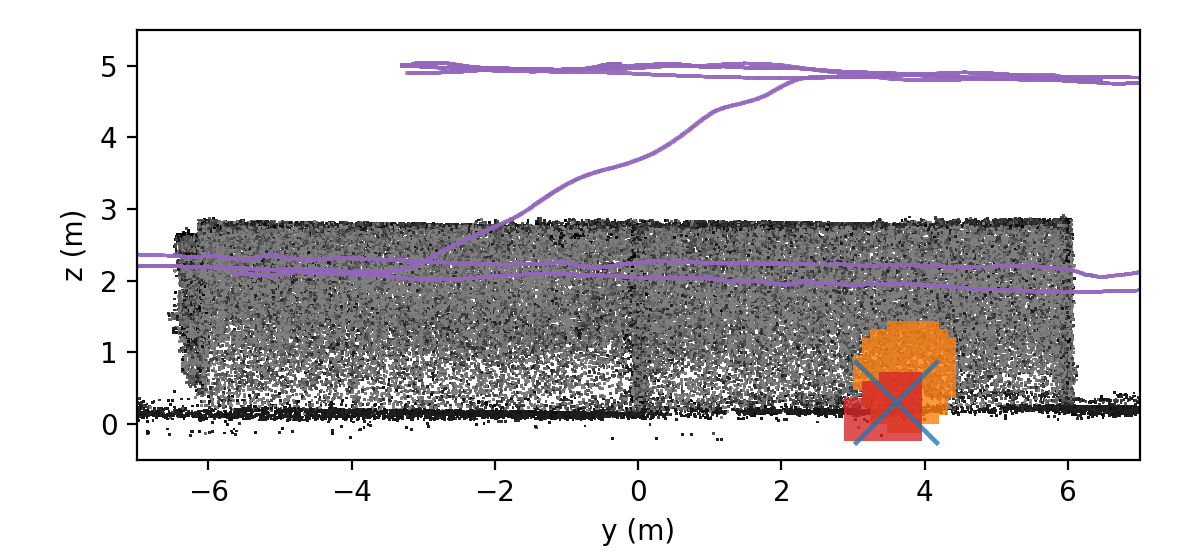}
\end{center}
\caption{Results for \isot{Ba}{133} showing the best fit model when the containers are assumed to be empty (orange) and when all voxels inside the top right container are marked as occupied (red).
The top plot shows the count rates, while the bottom two plots show the 2\textsigma~spatial confidence intervals for top-down and side views in the same colors as the count rate plot.
The ground truth is once again shown with a blue cross.
In this case, the result assuming the containers are empty coincidences with the result when solid attenuation is ignored, so that model and confidence intervals are not shown.\label{fig:ba133_results}}
\end{figure}

Inquiring further, it was discovered that the container adjacent to the empty container with the source was filled with various pieces of large scientific equipment, including many items made of steel.
So the same analysis was performed but marking all voxels inside that one container as occupied.
Once this was done, the log likelihood-MFP curve became nearly completely concave again, and the optimal MFP was found to be 2.58\,m with a 2\textsigma~confidence interval of 1.11--7.05\,m (see~\Fref{fig:nll_vs_mfp}).
Moreover, the 2\textsigma~confidence interval for the source activity shifted from 280--550 to 425--800\,\textmu Ci, which now includes the true activity (638\,\textmu Ci).
At the same time, the 2\textsigma~spatial confidence interval changes from excluding to including the ground truth (\Fref{fig:ba133_results}).

As with the \isot{Cs}{137} analysis, the best fit value of \(\lambda_{\mathrm{solid}}\) again is much larger than the tabulated value of 0.0133\,m for 356\,keV photons in steel~\cite{mcconn_jr_compendium_2011, xcom}, and this will be discussed in the next section.

\section{Discussion}\label{sec:discussion}
We have shown that attenuation by objects in the environment can be a major influence on the effectiveness of certain point source reconstructions, and it is possible to use three-dimensional information to estimate the effect of attenuation by material in the environment and obtain the correct location and activity of a point source, at least under certain conditions.
To demonstrate this result, experimental data were collected for two sources in shipping containers, and analysis yielded source parameters in agreement with the ground truth.

As noted earlier, the values of the best fit photon MFPs seem at first to be far too large when compared with tabulated values for attenuation by the steel walls of the containers --- 2.58 and 0.82\,m, versus 0.0133 and 0.0176\,m for \isot{Ba}{133} and \isot{Cs}{137}, respectively.
For \isot{Cs}{137}, this apparent discrepancy can be reconciled by the fact that the voxels were much thicker than the walls of the containers, which are approximately 2\,mm thick, although the corrugation of the steel increases the effective wall thickness to approximately 3\,mm~\cite{ling_technical_2020}.
In this case, the relevant parameter is not \(\lambda_{\mathrm{solid}}\) but the dimensionless quantity \(\lambda_{\mathrm{solid}} / \Delta\), the ratio of the MFP to the voxel pitch, which should be approximately equal to the true MFP (\(\lambda_{\mathrm{true}}\)) over the true characteristic wall thickness \(\Delta_{\mathrm{true}}\).
The expected ratio for 662\,keV photons is therefore 5.9, in agreement with the 2\textsigma~confidence interval 4.5--9.8.
Therefore, this scenario is believed to be explained primarily by steel attenuation.

On the other hand, the 2\textsigma~confidence interval of the \isot{Ba}{133} ratio \(\lambda_{\mathrm{solid}} / \Delta\) is 9.2--58.7, which excludes the tabulated value of 4.4 for the steel walls, but the analysis has already shown that there was attenuation from other materials in addition to the steel walls.
Also, the comparison of \(\lambda_{\mathrm{solid}} / \Delta\) to \(\lambda_{\mathrm{true}} / \Delta_{\mathrm{true}}\) is only valid when the attenuating material is thinner than the voxel size: \(\Delta_{\mathrm{true}} < \Delta\).
If the attenuating material is thicker than the voxel size, which is the case here for the bulk material inside the container, then the optimal MFP will increase to account for the thicker material: \(\lambda_{\mathrm{solid}} / \Delta \sim n_{\mathrm{voxels}} \cdot \lambda_{\mathrm{true}} / \Delta_{\mathrm{true}}\), where \(n_{\mathrm{voxels}} \sim \Delta_{\mathrm{true}} / \Delta\) is the number of voxels that the attenuating material spans on average.
For this particular measurement, the material does not uniformly fill the container and is of mixed contents, so no simple theoretical comparison can be made.

The analysis of measured data shows that PSL with attenuation estimation can work in at least some real-world scenarios, but it highlights at least one major challenge --- by relying on the LiDAR point cloud alone to sense the 3-D environment, the 3-D model is only capable representing surfaces, whereas the contents of the volumes contained within the exterior surfaces are unknown.
For some volumes, such as buildings with windows, the LiDAR is able to sense some of the interior and detect interior walls, but for many surfaces such sensing is impossible with LiDAR.
When attenuation from surfaces alone can explain most of the modulation of the source signal, then this method works well.
This situation was seen in the case where \isot{Cs}{137} was located in the gap between containers.
On the other hand, if attenuation cannot be explained by surfaces alone but by the material behind the surfaces, the method will not work as well until other assumptions are included, such as the \isot{Ba}{133} example.

In future work, it will be important to make inferences about the effect of unobserved volumes.
One approach could be to draw rays from the LiDAR during the measurement to determine which voxels in the model have not been visible to the LiDAR and thus are indeterminate, i.e., not known to be occupied or unoccupied.
PSL can then be performed by making some set of assumptions about the indeterminate voxels, such as introducing a second mean free path parameter and optimizing over it.
This approach would add yet another parameter to optimize over, with increased computational burden.

Additional contextual data may be helpful for this problem as well.
Visual cameras, combined with modern computer vision techniques, could label materials in the scene or provide some inference about the makeup of objects.
For example, knowing that a planar surface is a brick wall could allow one to make an assumption about its attenuation that will be robust for many photon energies without making any direct measurements of the wall's thickness.
Another example would be an algorithm identifying cargo containers in the scene, thus allowing an algorithm to posit guesses about what the volume's properties would be if it were empty or filled and to test those different possibilities.
A final example could take advantage of automated segmentation of 3-D volumes and a user interface that allows an operator to enter information based on domain knowledge.

Regarding fieldability, the implementation of PSL with attenuation estimation shown in this work does not run in real-time, and far from it (210~minutes for the \isot{Cs}{137} analysis and 100~minutes for the \isot{Ba}{133} analysis).
However, the processing here has been done on a single CPU, which is the worst case scenario.
In principle, the ray casting can be parallelized and run on GPUs, and a single PSL optimization has already been implemented on a GPU~\cite{hellfeld_free-moving_2021}.
A better optimization procedure than a brute force grid search and better initial guesses based on contextual information should also be able to speed up evaluation time, and constraints from computer vision classification or operator input could further reduce the computational complexity and afford nearly real-time inference.


\section{Acknowledgments}
The authors wish to thank Erika Suzuki for the use of her aerial photography.


\bibliographystyle{IEEEtran}
\bibliography{quant_atten_paper}

\end{document}